\documentclass[12pt]{article}

\usepackage[a4paper, margin=1in]{geometry}

\usepackage{color}

\usepackage{graphicx}
\usepackage{amsmath}
\usepackage{bm}
\usepackage{algorithm}
\usepackage{algpseudocode}

\algnewcommand{\LineComment}[1]{\State {\color{blue} \(\triangleright\) #1}}

\newcommand{\beginsupplement}{%
        \setcounter{table}{0}
        \renewcommand{\thetable}{S\arabic{table}}%
        \setcounter{figure}{0}
        \renewcommand{\thefigure}{S\arabic{figure}}%
     }

\newcommand{\de}[2]{\frac{d #1}{d #2}}

\newcommand{\figref}[1] {Figure \ref{#1}}
\newcommand{\tabref}[1] {Table \ref{#1}}

\definecolor{midgreen}{rgb}{0,0.5,0}

\title{Unlocking datasets by calibrating populations of models to data density: a study in atrial electrophysiology} 

\author{Brodie A. J. Lawson$^{1\ast}$, Christopher C. Drovandi$^{1}$, Nicole Cusimano$^{2}$, \\
Pamela Burrage$^{1}$, Blanca Rodriguez$^{3}$, Kevin Burrage$^{1,3\dag}$\\
\\
\normalsize{$^{1}$Australian Research Council Centre of Excellence for Mathematical and Statistical Frontiers, }\\
\normalsize{School of Mathematical Sciences, Queensland University of Technology, Brisbane, Australia}\\
\normalsize{$^{2}$Basque Center for Applied Mathematics, Bilbao, Spain}\\
\normalsize{$^{3}$Department of Computer Science, University of Oxford, Oxford, United Kingdom}\\
\\
\normalsize{$^\ast$To whom correspondence should be addressed; E-mail: b.lawson@qut.edu.au.} \\
\normalsize{$^{\dag}$ Visiting Professor}
}

\date{}

\begin{document} 




\maketitle 


\begin{abstract}
The understanding of complex physical or biological systems nearly always  requires  a characterisation of the variability that underpins these processes.  In addition, the data used to calibrate such models may also often exhibit considerable variability. A recent approach to deal with these issues has been to calibrate populations of models (POMs), that is multiple copies of a single mathematical model but with different parameter values. To date this calibration has been limited to selecting models that produce outputs that fall within the ranges of the dataset, ignoring any trends that might be present in the data. We present here a novel and general methodology for calibrating POMs to the distributions of a set of measured values in a dataset.  We demonstrate the benefits of our technique using a dataset from a cardiac atrial electrophysiology study based on the differences in atrial action potential readings between patients exhibiting sinus rhythm (SR) or chronic atrial fibrillation (cAF) and the Courtemanche–-Ramirez-–Nattel model for human atrial action potentials. Our approach accurately captures the variability inherent in the experimental population, and allows us to identify the differences underlying stratified data as well as the effects of drug block.
\end{abstract}


\section*{Introduction}

Mathematical modelling is vital for the understanding of complex phenomena, but the use of mathematical models requires careful specification of their parameter values against available data. In many applications, model predictions can vary sharply in response to even small changes in the values of their parameters, and yet experimental efforts to determine these values are invariably associated with either some kind of uncertainty or inherent variability underlying the processes that are being measured. In biological and physiological contexts, for example, not only are these uncertainties typically very large, the values of representative parameters also exhibit considerable variation between different members of a population, due to differences in physiology and genetics. Properly accounting for this variability using mathematical models is critical to furthering understanding in such fields \cite{Sarkar2012}.

With regards to uncertainty quantification (UQ), techniques such as Monte Carlo sampling \cite{Ogilvie1984}, polynomial chaos expansions \cite{Xiu2002}, and Bayesian approaches including Gaussian processes \cite{Ohagan1999}, allow for the impacts of uncertainty in parameter values upon model outputs (predictions) to be quantified, or for parameter values and their uncertainties to be determined in response to data collected for model outputs. However, each works from the perspective of a single, immutable model with some fixed uncertainties in its inputs and corresponding uncertainties in its outputs. This becomes an issue when one wishes for example to determine which features (parameter values) in a population predict different classes of outputs, or to consider the impacts of changes to the underlying model itself.

On the other hand, a very natural approach for modelling and understanding the variability within populations is the recent technique known as populations of models (POMs) \cite{Prinz2004, Marder2011, Britton2013}. In this approach, a collection of varying individuals is represented in kind by a collection of individual models, with the idea that the collection of models exhibits the same variability as the population being modelled. Although each individual model typically differs only in terms of the values of its parameters, each remains a model in its own right, allowing for subpopulations within the POM to be identified and analysed, and for the underlying model to be easily adjusted once a POM has been constructed. Somewhat related are genetic algorithms that use multiple copies of a model with differing parameter values as their organisms \cite{syed2005}, although there the focus is on breeding a single model that best fits data for a single individual, and not on characterising variability in a population.

Candidate models for a population can be generated simply by randomly sampling from a reasonable parameter space, but of course care should be taken to ensure that the models which compose the POM are physically (or physiologically) realistic. This is typically achieved by comparing the outputs of these candidate models to available experimental data, a process known as \textit{calibration} of POMs \cite{Muszkiewicz2016}. In the sphere of cardiac electrophysiology, where POMs research has been very active, calibrated POMs have been used to great effect. This includes suggesting modifications to existing models of rabbit ventricular cells \cite{Gemmell2014} and human atrial \mbox{cells} \cite{Muszkiewicz2014} required to reproduce specific data, determining the electrophysiological properties that lead to the dangerous phenomena of alternans \cite{Zhou2013, Zhou2016} and atrial fibrillation \cite{Liberos2016}, and characterising the sources of the differing function of failing hearts \cite{Walmsley2013}. The technique has also been used to explore the variable response of a population to drug-induced potassium channel block \cite{Britton2013, Drovandi2016}, to the onset of ischemia in rabbits \cite{Gemmell2016}, and the effects of hypertrophic cardiomyopathy \cite{Passini2016}. Most relevant to our work, calibrated POMs were used by Sanchez \textit{et al.} \cite{Sanchez2014} to explore the differences between patients exhibiting sinus rhythm (SR) or chronic atrial fibrillation (cAF), including identification of the impacts of cAF-induced remodelling by considering the differences in channel conductances between the POMs calibrated to the healthy and pathological datasets. 

To date, calibration of POMs has been achieved almost exclusively by rejecting any trialled models that produce outputs that correspond to measurable quantities falling outside the ranges of observations for those same quantities in the dataset \cite{Muszkiewicz2016}. This prevents any obviously unphysical models from being accepted into the population, but does not necessarily guarantee that the resulting POM directly corresponds to the data. Selection of models according to the ranges of values observed in the data creates a feasible region that is necessarily hyperrectangular, whereas the actual multidimensional spread of experimental measurements may be a much more complex shape. Additionally, it may be desirable that the selected models are not only \textit{feasible}, but also that together they exhibit the same features as seen in the data (such as regions of high or low density, and correlations between measured quantities). One recent work did calibrate a POM by ensuring that the models together exhibited appropriate mean and standard deviation for the output variable of interest \cite{Lancaster2016}, a step towards the distribution-driven calibration technique we introduce in this publication.

In this work, we extend a recent statistically-informed sampling technique for POM construction \cite{Drovandi2016} in order to propose a new method that produces POMs that are directly calibrated to data distributions. A consequence of this process is that it ensures that selected models lie in data-dense regions of the space of observables wherever possible. We demonstrate our technique using the data of Sanchez \textit{et al.}, taken from patients exhibiting either healthy SR of cAF. This allows us to show not only how our calibrated POMs serve as accurate \textit{in silico} representations of variable populations, but also how capturing specific features in data illuminates the differences between stratified populations including electrophysiological features that underpin the cAF pathology, as well as variable responses to drug block treatment. We finally conclude by discussing our new approach, when and how it should be used, and the implications for modelling and understanding variability in all its manifestations. 

\section*{Results}
\label{sec:results}

\subsection*{SMC Significantly Improves Calibration of Populations of Models to Data}
\label{sec:results_SMC}
The action of the heart depends on the excitable, highly nonlinear \cite{Qu2014} nature of cardiac cells, which undergo a carefully-controlled process of ion uptake and release in response to electrical stimulus. In addition to the temporary intake of Ca$^{2+}$ ions that produces the cellular contraction associated with the heartbeat, control of the potential difference across the cell's membrane also prevents it from being re-stimulated too quickly. The change in membrane potential in response to stimulus is known as the action potential (AP), and it is commonly recorded in single-cell experiments.

The Sanchez \textit{et al.} data quantifies APs recorded from the right atrial appendages of patients exhibiting either healthy SR or cAF in terms of their \textit{biomarkers}, measures of key AP properties that together define the important features of the AP's shape. These biomarkers are the action potential duration (APD), action potential amplitude (APA), resting membrane potential (RMP), maximum upstroke velocity ($\de{V}{t}_{\mbox{\scriptsize max}}$) and potential at 20\% repolarisation (V$_{20}$), with more information regarding the biomarkers provided in Materials and Methods. The biomarkers are the output variables of the model in our case, and so our goal is to construct populations of AP models that produce values of these biomarkers that exhibit the same variability as that seen in the experimentally recorded values. We use our sequential Monte Carlo (SMC) algorithm (detailed in Materials and Methods) to select models that, when taken together as a population, possess this property, and compare these to the POMs generated using Latin hypercube sampling (LHS) \cite{Mckay1979} matched to the ranges of the data, the typical approach for the construction of calibrated POMs \cite{Muszkiewicz2016}. We note that the SMC approach laid out by Drovandi \textit{et al.} \cite{Drovandi2016} could also be used to generate samples matched to biomarker ranges. Range-based calibration is highly appropriate in the case of a low number of experimental recordings, where there is insufficient information to derive an approximate distribution for the data.

POMs were constructed using the Courtemanche--Ramirez--Nattel (CRN) model \cite{Courtemanche1998} for atrial APs, with the conductance of all its currents important to the AP (eleven in total, including intracellular Ca$^{2+}$ uptake and release) allowed to vary by $\pm 100\%$ of the values originally published. Selection of a wide parameter space allows us to successfully find parameter values appropriate for the data against which we calibrate, though we also explore later the effects and implications of our calibration processes when we vary parameters by a smaller extent. Further information and justification regarding the choice of AP model and the currents that are varied is provided in Materials and Methods. Initialising the SMC algorithm with 2000 particles, we obtained POMs composed of 1938 unique models for the SR data (those exhibiting sinus rhythm), and 1931 unique models for the cAF data (those exhibiting chronic atrial fibrillation). Using 10000 trialled models generated using LHS (10 samples with 1000 divisions in each parameter dimension) produced 1319 accepted models for the SR dataset and 1338 models for the cAF dataset. We note that these numbers should not be compared as a measure of efficiency, for the SMC algorithm involves multiple model runs for each particle and is solving a more difficult sampling problem that takes the distributions of the data into account.

\begin{figure}
\begin{center}
\includegraphics[width=15cm, trim={4cm 2cm 4cm 1cm},clip]{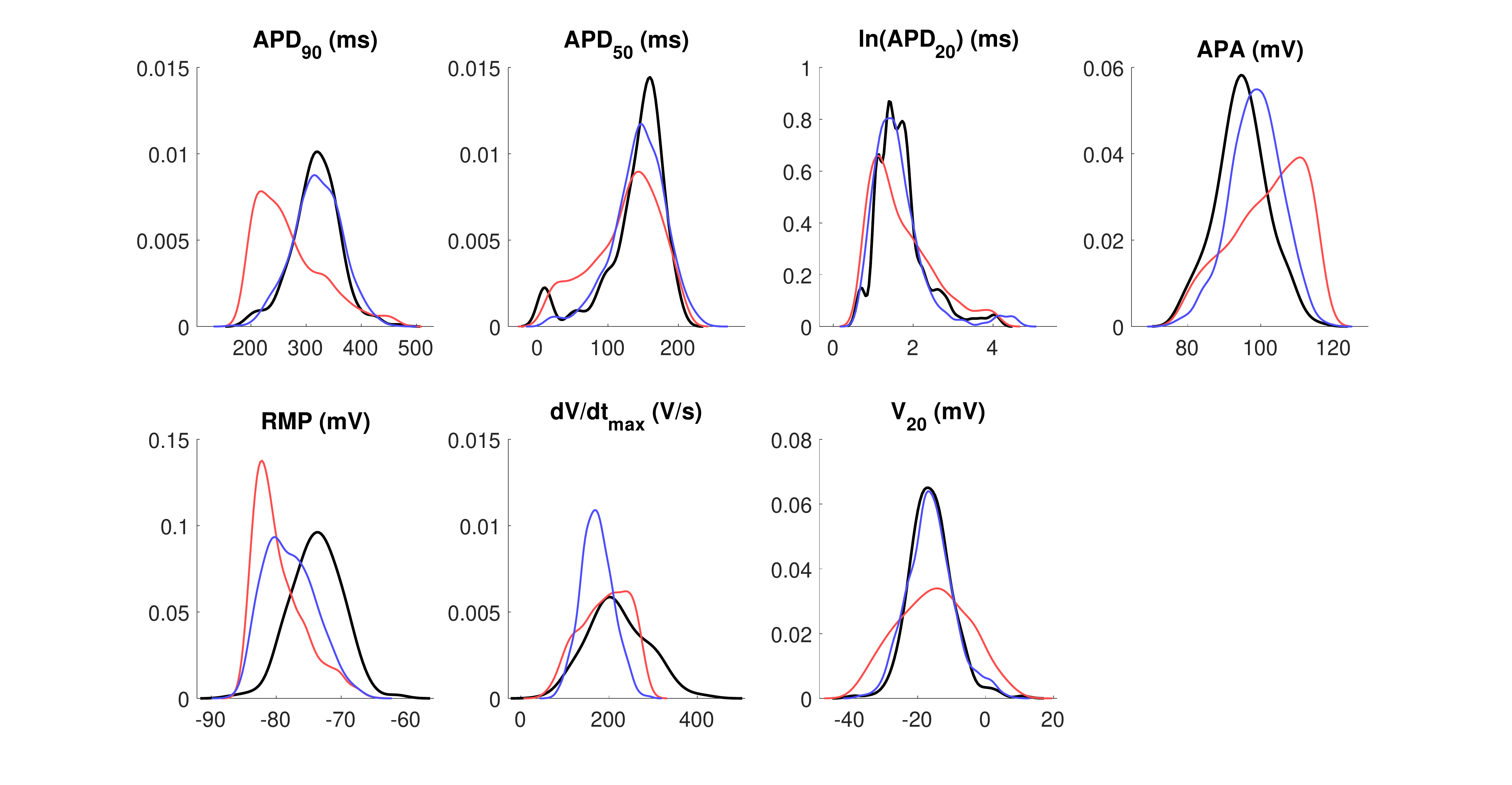}
\end{center}
\caption{{\bf{Calibration to biomarker distributions as opposed to their ranges  reduces model bias.}} Marginal distributions of the biomarkers in the SR dataset (black) and POMs calibrated to biomarker distributions using the SMC algorithm (blue) or calibrated to biomarker ranges using LHS (red). The natural logarithm of APD$_{20}$ values is used to better display their distribution.}
\label{fig:SMCvsLHS_marginals}
\end{figure}

For the SR dataset the SMC-calibrated POMs show a significantly better degree of localisation to data-dense regions in the biomarker space, when compared to POMs calibrated to biomarker ranges (\figref{fig:SMCvsLHS_marginals}). The effects of bias inherent to the underlying model are greatly reduced, as best demonstrated by the biomarkers APD$_{90}$, RMP and APA. Over the selected range of parameters, the CRN model tends to produce APD$_{90}$ values lower than the majority of the data, and APA values higher than the majority of the data, and when matching to ranges there is nothing to prevent these models from being over-represented in the final population. The SMC-constructed population captures very well the extent of variance in the V$_{20}$ values in the dataset. However, bias is still clearly present in RMP and $\de{V}{t}_{\mbox{\scriptsize max}}$.  Very similar results are seen for the cAF dataset (see Figures \ref{fig:SMCvsLHS_CA_marginals} and \ref{fig:SMCvsLHS_CA_biopairs}).

\begin{figure}
\begin{center}
\includegraphics[width=15cm, trim={4cm 2cm 4cm 1cm},clip]{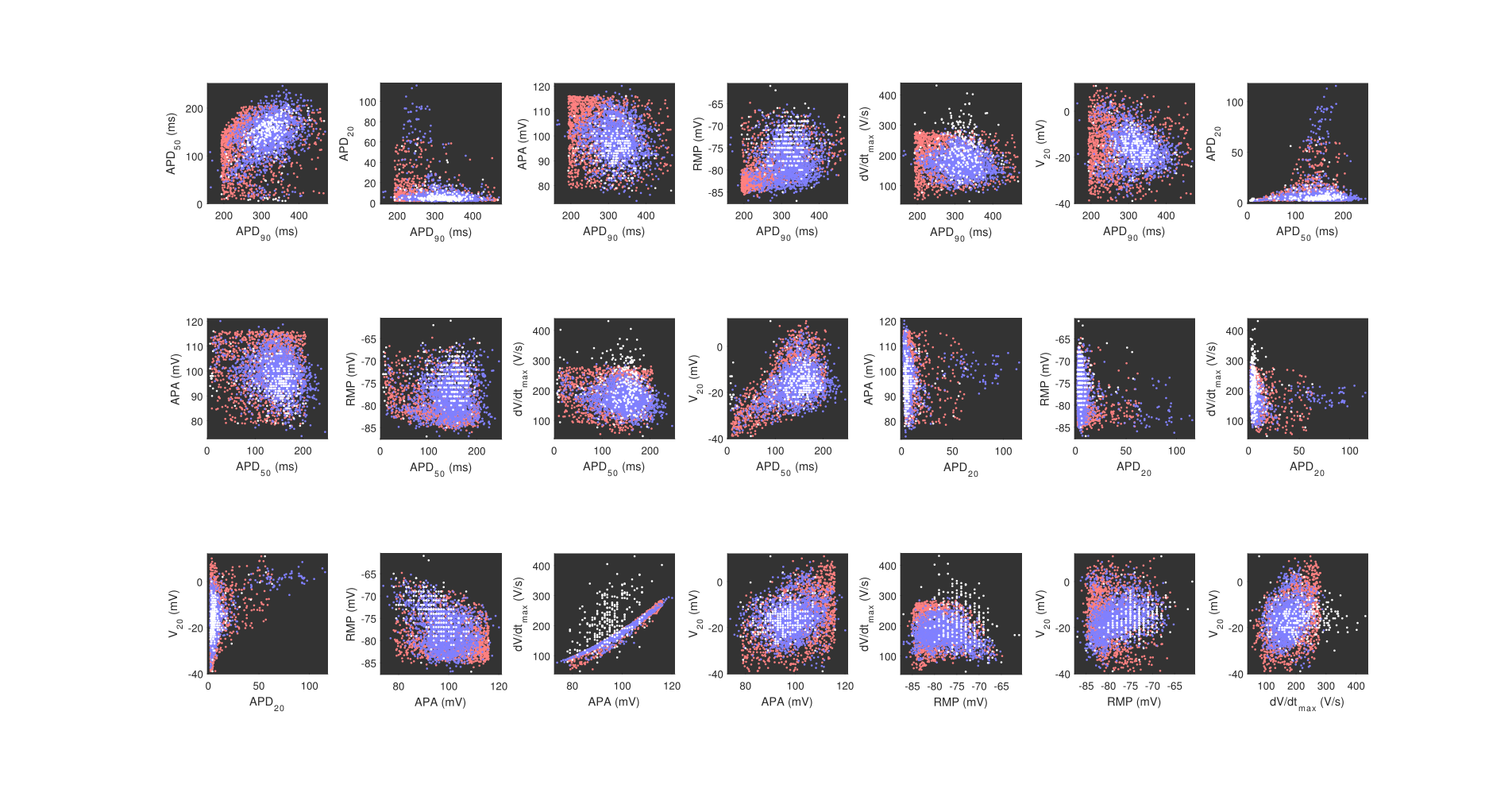}
\end{center}
\caption{{\bf{Bivariate distributions of biomarker pairs are well captured by an SMC-constructed POM.}} Pairwise scatterplots of each unique pair of biomarkers in the SR dataset (white) and the POMs constructed using SMC matched to distributions (blue) and LHS matched to ranges (red). The SMC-generated POM demonstrates good localisation to the dense regions in the data, but clearly requires further calibration. An obvious correlation between APA and $\de{V}{t}_{\mbox{\scriptsize max}}$ is exhibited by the model, regardless of the sampling method used, but this correlation is not present in the data.}
\label{fig:SMCvsLHS_biopairs}
\end{figure}

It is clear that the CRN model cannot produce certain combinations of biomarker values, regardless of the parameter values chosen within the search space, as can be seen when  different pairs of biomarkers are plotted (\figref{fig:SMCvsLHS_biopairs}). For example, the CRN model demonstrates a very clear correlation between the maximum upstroke velocity and the AP amplitude, a relationship that is not present in the data. We suggest this is not a failing of the CRN model (indeed, it is expected that a faster upstroke will allow for a higher peak membrane potential to be achieved while the cell's ion channels adjust), and thus the discrepancy between model and data likely arises due to the difficulties in measuring the maximum upstroke velocity accurately in an experimental setting. We also note that the POMs constructed here all use a single consistent stimulus protocol for all members of the population, whereas experimental data collection is expected to show some variation in applied stimulus current between cells, presenting another potential source for this discrepancy. Nevertheless, the SMC algorithm successfully reduces the impacts of model bias on POM construction,  producing POMs that accurately reflect the features of the data.

\subsection*{Refinement via Selection of Optimal Subpopulations Captures Well the Variability in Datasets}
\label{sec:results_refinement}
When considering Figures \ref{fig:SMCvsLHS_marginals} and \ref{fig:SMCvsLHS_biopairs}, it is clear that there still remain some differences in distribution between the models selected by the SMC algorithm and the data, particularly in the case of the biomarker RMP. In order to address this issue, we include a second phase of our calibration process that selectively removes models from the SMC-constructed POMs in order to improve the correspondence between POM and data, which we term ``refinement''. The advantages of this refinement process in more closely matching the data distributions is made clear in the results that follow.

Our easily-approximated divergence measure $\rho$ provides a means for selecting this subpopulation, using a simulated annealing-type algorithm (see Materials and Methods, ``Further POMs Refinement''). Of course, selecting a subpopulation from a POM cannot produce coverage in areas of the biomarker space where the original POM has no models, and so the aforementioned issues with the maximum upstroke velocity cannot be addressed using our refinement technique. This motivates the use of a second divergence measure $\hat{\rho}$ (see Materials and Methods) that reduces the emphasis on this biomarker. We consider here the POMs formed by selecting subpopulations that minimise both measures.

\begin{figure}
\begin{center}
\includegraphics[width=15cm, trim={4cm 2cm 4cm 1cm},clip]{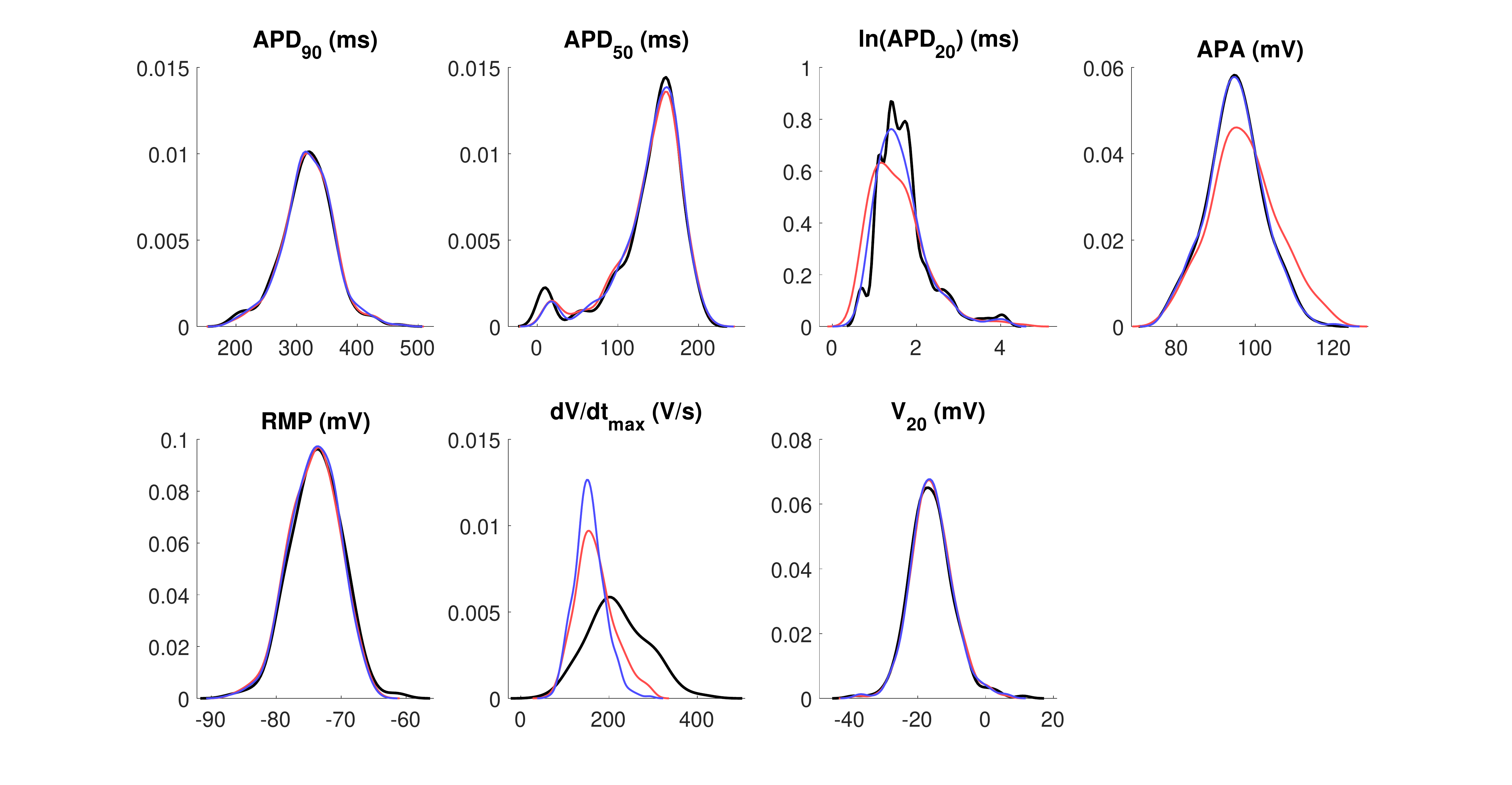}
\end{center}
\caption{{\bf{Selection of an optimal subpopulation almost fully captures biomarker variability.}} Marginal distributions of the biomarkers in the SR dataset (black) and POMs selected as subpopulations of the SMC-generated POM that minimised $\rho$ (red) or $\hat{\rho}$ (blue). The simulated annealing algorithm clearly succeeds at selecting a representative subpopulation, but the distributions of the APA and $\de{V}{t}{\mbox{\scriptsize max}}$ are not quite captured. Reducing the emphasis of $\de{V}{t}{\mbox{\scriptsize max}}$ on the calibration process provides very good capture of variability in all other biomarkers.}
\label{fig:SR_marginals}
\end{figure}

For the SR dataset, minimising $\rho$ produced a refined POM of 275 models with very good representation of variability in the dataset (\figref{fig:SR_marginals}), with the marginal distributions of the biomarkers matching very well apart from APD$_{20}$, APA and $\de{V}{t}_{\mbox{\scriptsize max}}$. The already observed coupling of APA and $\de{V}{t}_{\mbox{\scriptsize max}}$ makes it impossible to find models that simultaneously capture the distributions of these two biomarkers, and we attempt to rectify this using the modified divergence measure. By reducing the emphasis on $\de{V}{t}_{\mbox{\scriptsize max}}$, the refinement process is free to select models which at least capture the distribution of APA values, the biomarker we expect to be more accurately measured. 

When a subpopulation that minimises $\hat{\rho}$ is selected, the result is a POM composed of 327 models that shows slight improvements in the marginal distributions of the other biomarkers and a more significant improvement in  APD$_{20}$ (\figref{fig:SR_marginals}). Most notably, however, the distribution of APA values is now exceedingly well-represented in the POM, as the algorithm's efforts to minimise the divergence from the maximum upstroke velocity no longer hamper its ability to fit the distribution of the highly correlated APA. This improved performance also comes at little cost to the value of the unmodified divergence measure $\rho$, as shown by \tabref{tab:rhos}, and so we favour POMs constructed by minimising $\hat{\rho}$ in the remainder of this work.

\begin{table}[ht]
\begin{center}
\begin{tabular}{lcccccccc}
\hline
\bf{Population of Models} & & \multicolumn{3}{c}{\bf{SR data}} & & \multicolumn{3}{c}{\bf{cAF data}}\\
 & & $\rho$ & & $\hat{\rho}$ & & $\rho$ & & $\hat{\rho}$ \\ \hline
LHS, matched to ranges & & 2.76 & & 2.41 & & 2.74 & & 2.42 \\
SMC, matched to distributions & & 1.95 & & 1.41 & & 2.16 & & 1.74 \\
SMC subpopulation, minimising $\rho$ & & 1.36 & & 0.68 & & 1.21 & & 0.70 \\
SMC subpopulation, minimising $\hat{\rho}$ & & 1.47 & & 0.49 & & 1.28 & & 0.57 \\
\hline
\end{tabular}
\end{center}
\caption{{\bf{SMC produces more representative POMs than LHS matched to ranges for both datasets, and overall performance after subsequent refinement is similar in both cases.}}  
Comparison of the ability of different POMs to capture the between-subject variability in two clinical datasets, as provided by the divergence measures $\rho$ and $\hat{\rho}$. Lower $\rho$ values indicate a better fit to the distributions, demonstrating a significant gain from both the  SMC and from choosing an optimal subpopulation. For reference, a complete divergence between data and POM would produce $\rho_{\mbox{\scriptsize max}} = 5.83$ and $\hat{\rho}_{\mbox{\scriptsize max}} = 5.00$.}
\label{tab:rhos}
\end{table}

\begin{figure}
\begin{center}
\includegraphics[width=15cm, trim={4cm 2cm 4cm 1cm},clip]{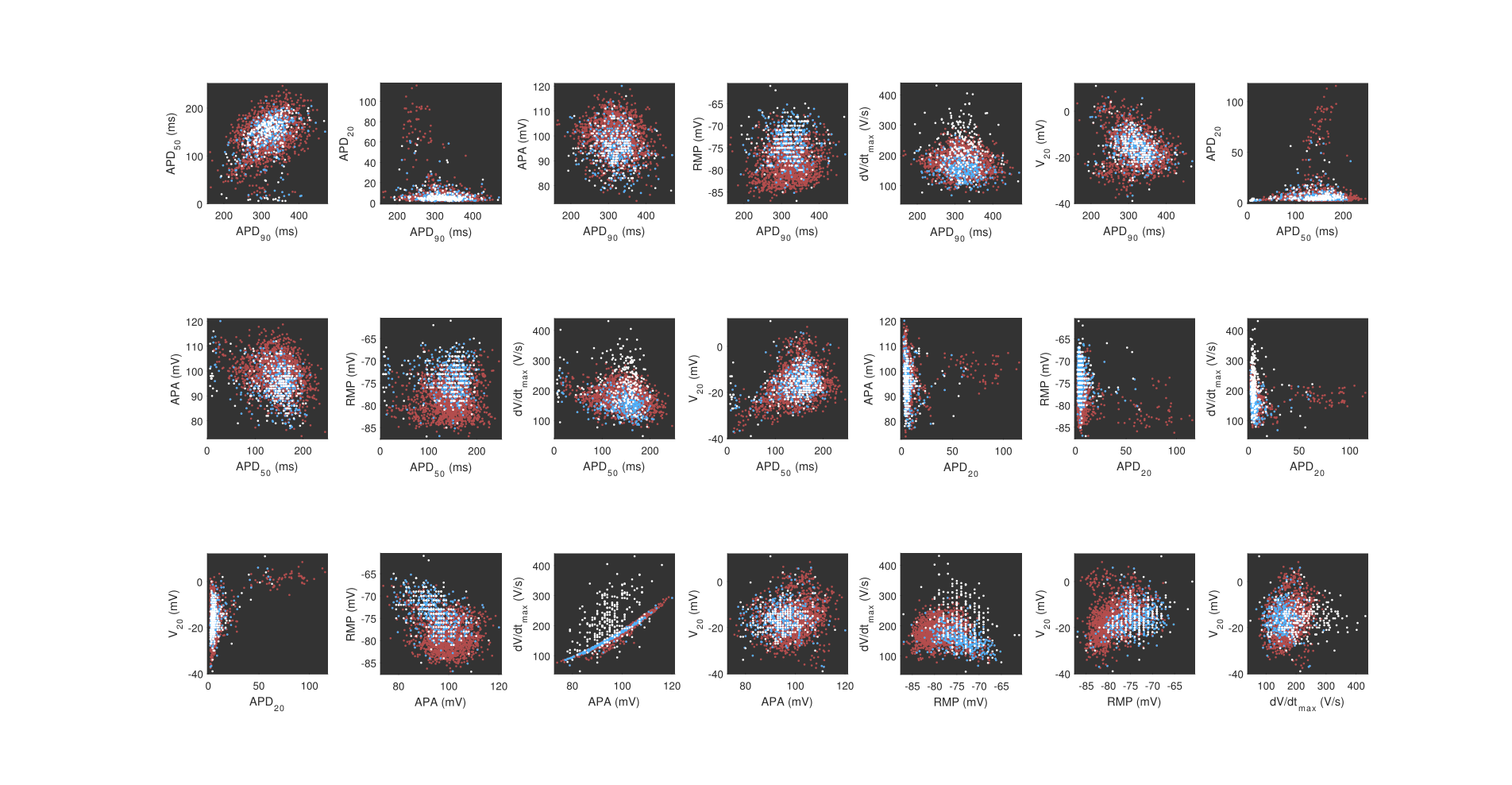}
\end{center}
\caption{{\bf{Simulated annealing successfully selects models according to data density in the biomarker space.}} Pairwise scatterplots of each unique pair of biomarkers in the SR dataset (white) and the models from the SMC-generated POM that were accepted (light blue) or rejected (dark red) in the process of minimising $\hat{\rho}$. Outside of $\de{V}{t}_{\mbox{\scriptsize max}}$, the features of the data are very well represented by the final POM.}
\label{fig:SR_biopairs}
\end{figure}

Our refined POM captures the key statistical properties of each of the biomarkers in the dataset very well, except for $\de{V}{t}_{\mbox{\scriptsize max}}$ which still suffers from the model's general underestimation of this value compared to the data ( \tabref{tab:SR_summaries}). \figref{fig:SR_biopairs} provides a visual demonstration of the refinement process, showing the selection of a subpopulation of models that corresponds well to the density of data across the biomarker space. Similarly good performance is also achieved by refined POMs calibrated to the cAF dataset (Figures \ref{fig:CA_marginals} and \ref{fig:CA_biopairs} and \tabref{tab:CA_summaries}), with again only $\de{V}{t}_{\mbox{\scriptsize max}}$ showing any significant deviation between POM and data.

\begin{table}[ht]
\begin{center}
\begin{tabular}{lllll}
\hline
\bf{Biomarker}  & \bf{Range} & \bf{Mean} & \bf{Std. Dev.} & \bf{JSD} \\ \hline
APD$_{90}$ (ms) & 191$ - $470 (193$ - $467) & 319 (318) & 43 (44) & 0.026 \\
APD$_{50}$ (ms) & 7$ - $215 (6$ - $206) & 142 (139) & 40 (44) & 0.066 \\
APD$_{20}$ (ms) & 2$ - $61 (2$ - $63) & 7 (7) & 7 (8) &  0.143 \\
APA (mV) & 77$ - $120 (78$ - $116) & 95 (95) & 7 (7) &  0.021 \\
RMP (mV) & -87$ - $-65 (-87$ - $-61) & -74 (-74) & 4 (4) & 0.059  \\
V$_{20}$ (mV) & -37$ - $6 (-39$ - $11) & -16 (-16) & 6 (6) & 0.041 \\
$\de{V}{t}_{\mbox{\scriptsize max}}$ (V/s) & 68$ - $292 (48 $ - $431) & 156 (220) & 34 (68) & 0.426  \\ \hline
\end{tabular}
\end{center}
\caption{{\bf{Summary statistics for the SR dataset are well recovered by the calibrated POM.}} Summary statistics for the POM calibrated to the distributions obtained by minimising $\hat{\rho}$ in biomarkers exhibited by atrial cells in sinus rhythm, as compared to the summary statistics for the experimental data itself (given in parentheses). Deviation in the marginal distributions of each biomarker are specified in terms of the Jensen-Shannon distance, calculated using equation (\ref{JSD}). The statistical variation of the data is seen to be well captured, apart from the maximum upstroke velocity, which also manifests in a large JSD value.}
\label{tab:SR_summaries}
\end{table}

Our two-phase calibration technique is thus seen to be successful at producing POMs that accurately reflect the data, at least to the extent that the underlying model can over the specified parameter space. Here this has been achieved for multiple datasets that show significant variation between individual samples, using a strongly nonlinear model and a moderately high number of variable parameters. Although the selection of optimal subpopulations did require the rejection of a large proportion of the models selected by the SMC algorithm in this case ($\sim$83\% for the SR dataset), larger POMs can be generated as desired by using additional particles in the original SMC algorithm, or by modifying the `energy' minimised in the refinement process so that POMs of a larger size are encouraged.

\subsection*{Distributional Calibration Produces  Populations of Models that Capture Important Data Features}

\begin{figure}
\begin{center}
\includegraphics[width=12cm, trim={4cm 0.5cm 4cm 1cm},clip]{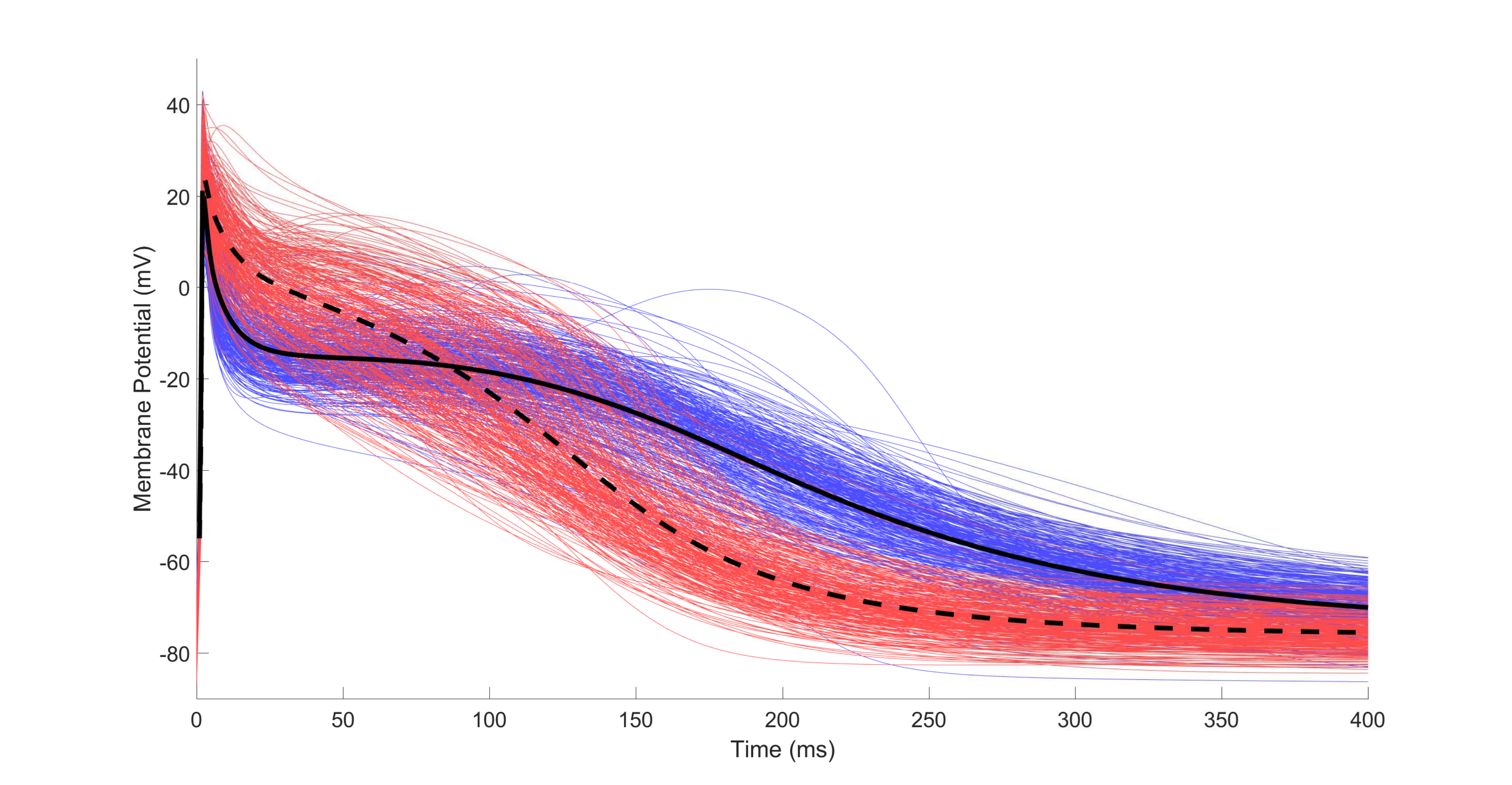}
\end{center}
\caption{{\bf{Distributional calibration captures the morphological differences between SR and cAF atrial action potentials.}} Atrial action potentials produced by simulation of the populations of CRN models calibrated to biomarker data for patients exhibiting sinus rhythm (blue) and chronic atrial fibrillation (red). Also displayed are the average of all traces for the sinus rhythm (solid) and atrial fibrillation (dashed) populations. The increase in AP triangulation and reduced refractory period associated with chronic atrial fibrillation is clearly demonstrated, especially by the averaged traces.}
\label{fig:AP_traces}
\end{figure}

The POMs constructed for the SR and cAF dataset using our calibration technique  capture very well the features of the data, but a natural question is whether this calibration actually produces quantifiable differences in the models that are selected, and their outputs. Here, the most critical output of the population of CRN models are the APs,  shown in \figref{fig:AP_traces}. Significant differences between the SR and cAF populations are immediately observed. SR APs demonstrate an initial period of very rapid repolarisation after the AP peak, then an extended plateau phase that eventually resumes gradual repolarisation back to the resting potential. In contrast, cAF APs show much less significant initial repolarisation, but their lack of any significant plateau phase and overall faster repolarisation produces significantly lower APDs (and hence a decreased refractory period).

These features are well known to be associated with cAF, which is characterised by far more triangular APs that lack a noticeable plateau phase and return to resting potential more rapidly than APs in healthy sinus rhythm \cite{Boutjdir1986}. The morphological differences between SR and cAF APs are seen in the data, with the rapid repolarisation followed by plateau in the SR population implied by very small APD$_{20}$ values and larger APD$_{50}$ values, while the AP triangulation in the cAF population is seen in larger APD$_{20}$ but smaller APD$_{50}$ values. Calibrating to distributions naturally takes these features of the data into account, successfully selecting models that predict the appropriate morphologies. When calibrating to ranges for this dataset, the models selected do successfully produce APs that show the reduced APD as associated with cAF, but are less successful in predicting the accompanying differences in AP morphology (see \figref{fig:AP_traces_LHS}, also Figures 2 and 3 in \cite{Sanchez2014}). We note that previous studies have created additional biomarkers that can be expressed in terms of the original biomarkers, such as measures of triangulation based on combinations of different APD values, in order to allow more effective calibration by capturing these additional features --- see \cite{Walmsley2013}. Our method, however, does not depend upon identifying the important trends in data or designing additional outputs to capture them, making it generally applicable.

Accurate prediction of the specific shapes of the APs for SR and cAF patients is important as it suggests that the differential actions of the many ionic currents that together produce the AP are being well captured by the POMs calibrated to distributions. This is critical when it comes to using these POMs for further analysis, such as considering the response of the different members of the population to drug treatments that act on specific cellular currents \cite{Britton2013, Drovandi2016}, or identifying the differences in underlying electrophysiology that characterises the two populations. We demonstrate these aspects in the following subsection.

\subsection*{Distributional Calibration Produces POMs that Capture Key Atrial Electrophysiological Aspects}

\subsubsection*{(i) Impacts of cAF-induced Remodelling}

We have constructed POMs calibrated to the SR and cAF datasets by varying the relative strengths of the different currents that contribute to the human atrial AP, and thus any significant differences in parameter values selected for the two datasets suggest that it is changes in these currents that produce the modified APs associated with the cAF pathology. Indeed, electrical remodelling of atrial myocytes that changes the densities of their different ion channels is a well-known feature of cAF and contributes to the persistence of the condition \cite{Nattel2008}. Dobrev and Ravens \cite{Dobrev2003} provide a review of the experimental evidence for the changes in current density associated with cAF, although further remodelling has since been experimentally identified \cite{Pandit2013}.

\begin{figure}
\begin{center}
\includegraphics[width=12cm, trim={4cm 0.5cm 4cm 1cm},clip]{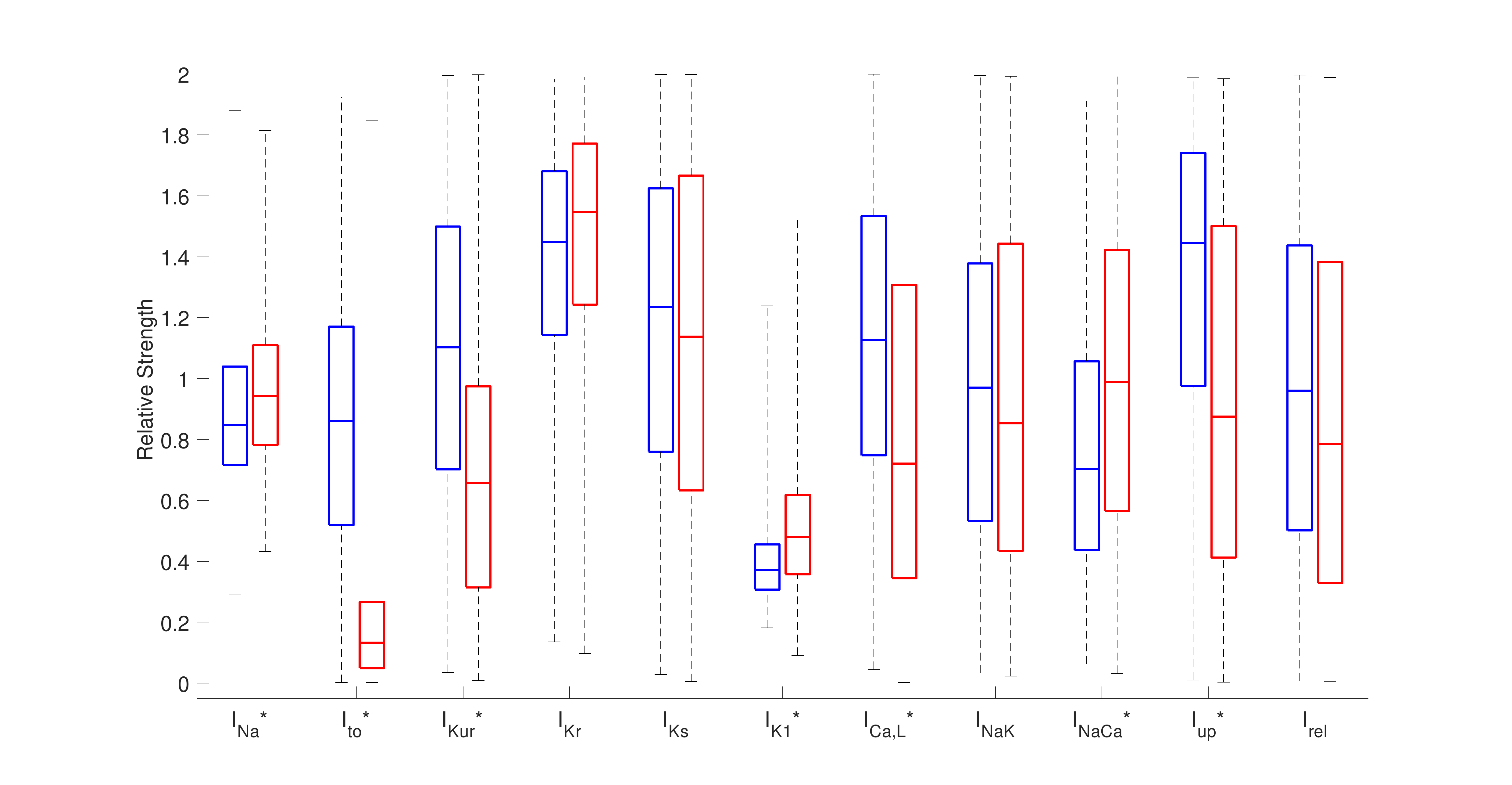}
\end{center}
\caption{{\bf{Accepted parameter values for the SR and cAF populations predict well the changes in ionic behaviour associated with the cAF pathology.}} Boxplot of $\boldsymbol{\theta}$ values composing the POMs calibrated to the SR (blue) and cAF (red) datasets. Values are expressed in relation to the base parameter values for the CRN model. Current densities that show statistically significant differences ($p < 0.001$ from the Mann-Whitney U test) are indicated with a *. The currents most well-known as remodelled in cAF ($I_{to}$, $I_{Kur}$, $I_{K1}$ and $I_{CaL}$) all show significant differences in the correct directions.}
\label{fig:theta_comparisons}
\end{figure}

Sanchez \textit{et al.} \cite{Sanchez2014} also compared the POMs generated for the CRN model and the models of Maleckar \textit{et al.} \cite{Maleckar2009} and Grandi \textit{et al.} \cite{Grandi2011} when calibrated to the ranges of the SR and the cAF data, by varing the six currents identified as most important to AP properties. They identified a statistically significant upregulation of $I_{K1}$ in all three models, with changes in other currents found to be model-dependent. In the case of the CRN model, Sanchez \textit{et al.} also found statistically significant decreases in $I_{CaL}$ and $I_{to}$ in accordance with experimental observation \cite{Nattel2008}. However, the observed decreases in $I_{to}$ and $I_{CaL}$ were quite small and the POMs constructed failed to identify $I_{Kur}$ as a significantly downregulated current in cAF. Furthermore, $I_{NaCa}$ showed a statistically significant decrease, despite Na$^{+}$/Ca$^{2+}$ exchanger action being known to \textit{increase} in cAF-afflicted atria \cite{Neef2010}. Our simulations using LHS calibrated to biomarker ranges also show an erroneous decrease in $I_{NaCa}$ and little to no change in $I_{CaL}$ and $I_{Kur}$, even with a larger suite of currents now allowed to vary. In contrast, SMC-calibrated POMs show significant differences in many current strengths between the SR and cAF populations (\figref{fig:theta_comparisons}), and do very well at identifying the currents that are known to be remodelled in response to cAF. We summarise the results in \tabref{tab:ATR}, comparing the experimentally observed changes in current density to those predicted by our POMs calibrated to full distributions or to only the ranges of the data.

\begin{table}
\begin{center}
\begin{tabular}{cccc}
\hline
\bf{Parameter} &  \bf{Exp.} & \bf{POMs} & \bf{POMs} \\ 
 & & \bf{(dist.)} & \bf{(ranges)} \\ \hline
$g_{Na}$ & $\updownarrow^{1}$ \cite{Sossalla2010} & $+11\%$ & $\leftrightarrow$ \\
$g_{to}$ & $\sim -70\%$ \cite{Dobrev2003} & $-85\%$ & $-51\%$\\
$g_{Kur}$ & $\sim -50\%$ \cite{Dobrev2003} & $-40\%$ & $-6\%$ \\
$g_{Kr}$ & $\leftrightarrow^{2}$ \cite{Pandit2013} & $\leftrightarrow$ & $+10\%$ \\
$g_{Ks}$ & $\sim$ +100\% \cite{Caballero2010} & $\leftrightarrow$ & $\leftrightarrow$ \\
$g_{K1}$ & $\sim +100\%$ \cite{Dobrev2003} & $+29\%$ & $+33\%$ \\
$g_{CaL}$ & $\sim -70\%$ \cite{Dobrev2003,Gaborit2005} & $-36\%$ & $\leftrightarrow$ \\
$I_{NaK\mbox{\scriptsize(max)}}$ & $\leftrightarrow$ \cite{Workman2003} & $\leftrightarrow$ & $+10\%$ \\
$I_{NaCa\mbox{\scriptsize(max)}}$ & $\sim +40\%$ \cite{Neef2010} & $+41\%$ & $-18\%$ \\
$I_{up\mbox{\scriptsize(max)}}$ & $\updownarrow^{3}$ \cite{ElArmouche2006,Voigt2009} & $-39\%$ & $\leftrightarrow$ \\
$k_{rel}$ & $\uparrow^{4}$ \cite{ElArmouche2006} & $\leftrightarrow$ & $\leftrightarrow$ \\
\hline
\end{tabular}
\end{center}
\caption{{\bf{Experimentally observed changes in current density associated with cAF are well predicted by POMs calibrated to distributions.}} Changes in median current activities between the POMs calibrated to either the distributions, or the ranges of the SR and cAF datasets, as compared with experimentally observed (Exp.) measurements of changes in current densities associated with this pathology. Experimental figures are taken from the specified references and rounded to the closest 10\% to reflect the general uncertainty in their measurements, and in some cases represent the combined result of multiple studies. The $\leftrightarrow$ symbol indicates no significant change observed ($p \geq 0.01$ from the Mann-Whitney U test for POMs), and other symbols used are explained by the following notes: $^{1}$Peak $I_{Na}$ current is reduced, but sustained $I_{Na}$ increased. $^{2}$Decreases in mRNA levels have been observed \cite{Dobrev2003}, but no direct experimental evidence for $I_{Kr}$ change in cAF has yet been provided. $^{3}$Ca$^{2+}$ uptake is reduced by decreased Serca2a levels, but increased by enhanced phosphorylation of SERCA inhibitors. $^{4}$Ca$^{2+}$ release is increased, but in a `leaky' fashion not necessarily best represented by changes to $k_{rel}$ in the CRN model. Distribution-calibrated POMs detect more of the differences in current densities that underlie the cAF pathology, correlating well with experimentally-observed changes in the greatest majority of current densities.}
\label{tab:ATR}
\end{table}

In analysing \tabref{tab:ATR} we see that the POMs calibrated to distributions underestimate the extent of upregulation of $I_{K1}$ and downregulation of $I_{CaL}$, but identify that these currents are, respectively, increased and decreased in cAF. The other critical current changes associated with cAF are the reductions in $I_{to}$ and $I_{Kur}$ \cite{Nattel2008}, that are also well detected. We also predict the increase in $I_{NaCa}$ activity that is expected. Although recent experimental evidence suggests a strong upregulation of $I_{Ks}$ in cAF \cite{Caballero2010}, which has not been observed here, this current has only a minor contribution to repolarisation in the CRN model and this is likely the reason for a lack of any significant difference in the $g_{Ks}$ values selected for the two populations. The results here suggest that the net effect of changes to atrial SERCA function (calcium uptake) is a decrease, as also used by Grandi \textit{et al.} to represent the cAF case in their AP model \cite{Grandi2011}. 

\subsubsection*{(ii) Response to Anti-arrhythmic Treatment}

 Arrhythmias in the heart are typically treated by drugs that block specific ion channels, reducing the impact of the corresponding current(s) on the action potential. A common target is the rapid component of the delayed K$^{+}$ rectifier current ($I_{Kr}$), which activates comparatively late in the AP and is a primary contributor to repolarisation in this phase. Reducing flow due to this current hence prolongs the AP, and can restore SR in patients with cAF \cite{Colatsky1994}. $I_{Kr}$ was also the current chosen to explore the differential response of a variable population to drug treatment using POMs in a previous study \cite{Britton2013}. 

\begin{figure}
a)\begin{center}
\includegraphics[width=9cm, trim={4cm 0.5cm 4cm 1cm},clip]{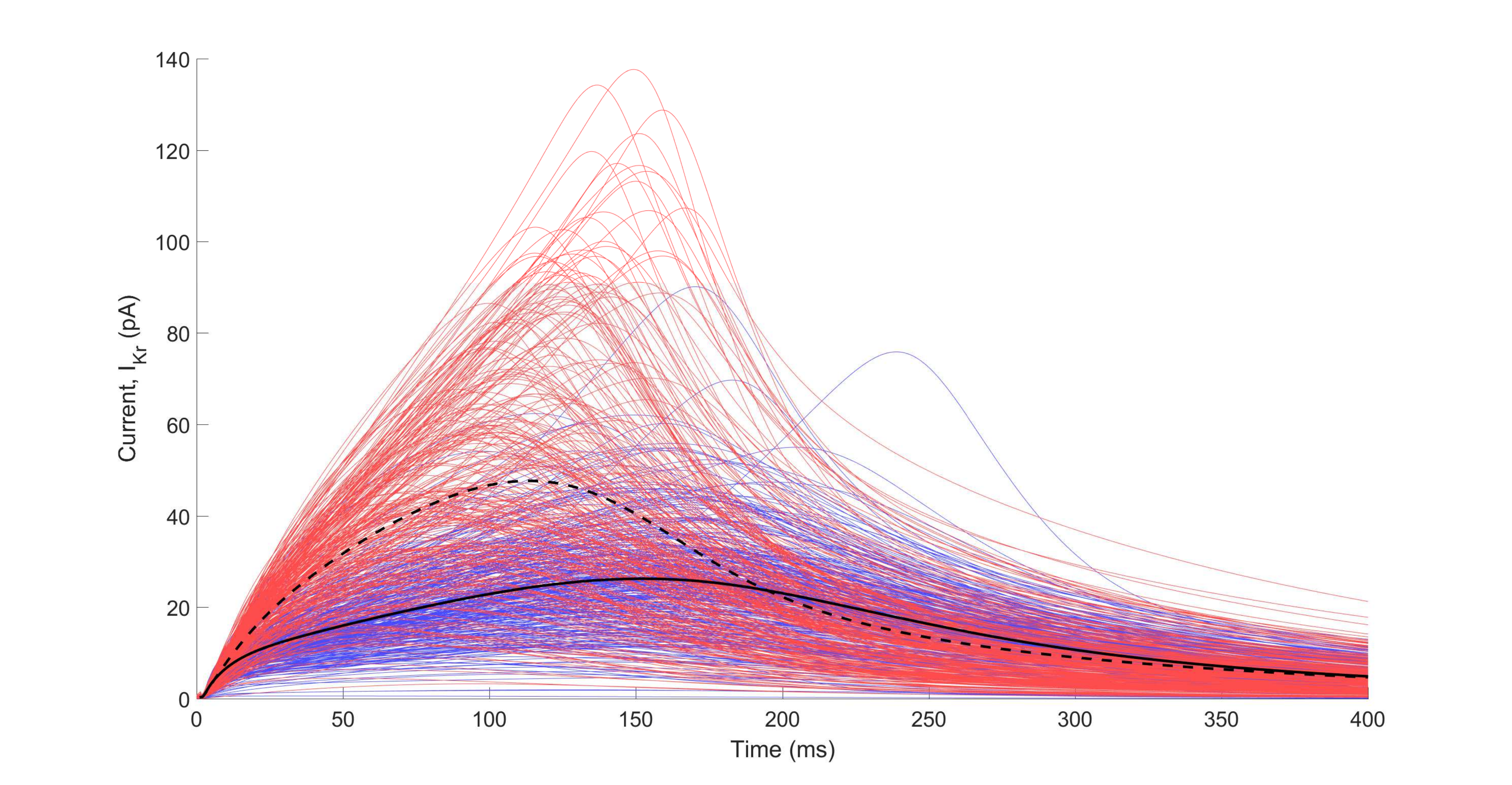}
\end{center}b)\begin{center}
\includegraphics[width=9cm, trim={4cm 0.5cm 4cm 1cm},clip]{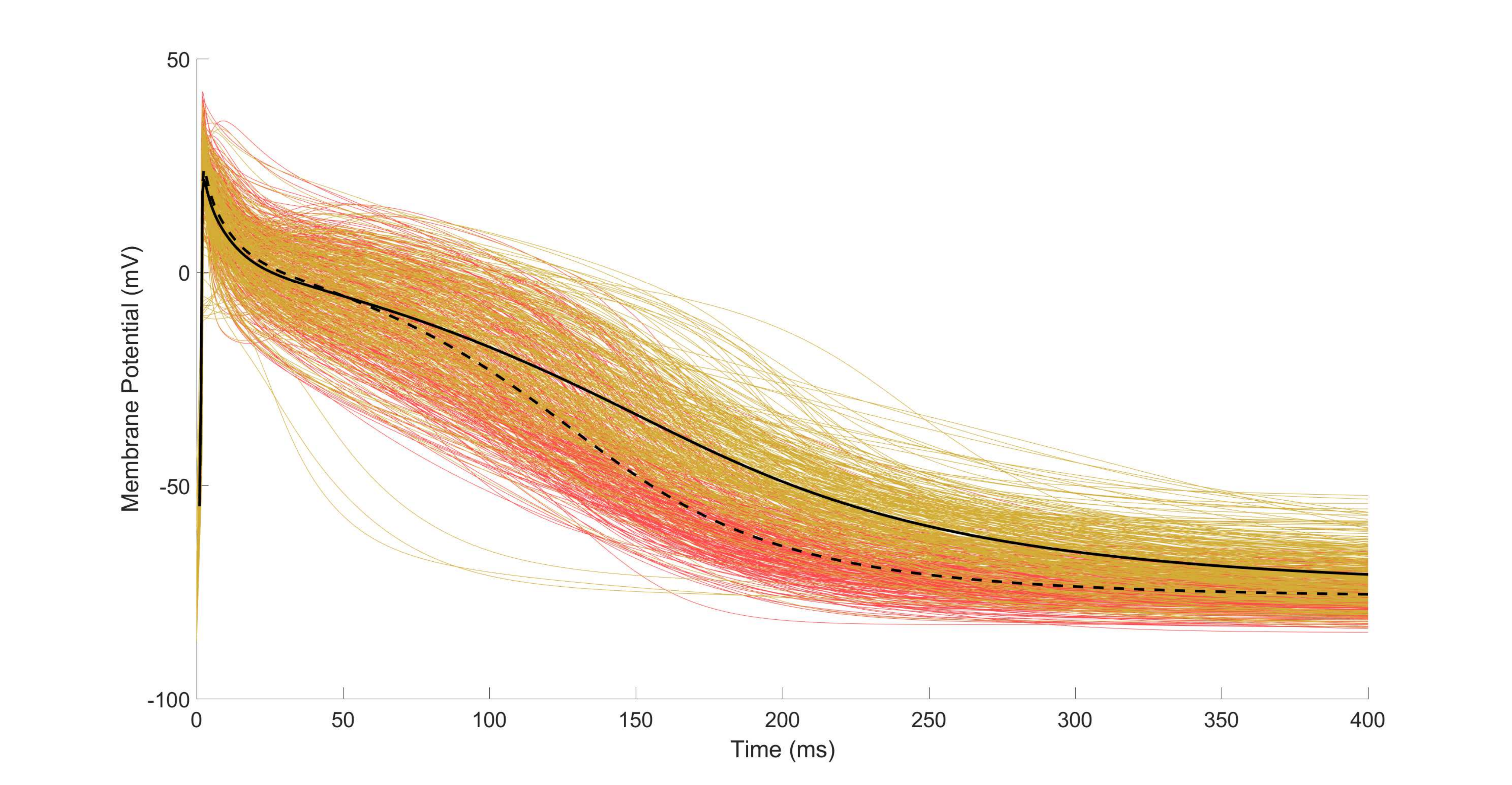}
\end{center}
\caption{{\bf{Calibration to distributions produces models that respond appropriately to anti-arrhythmic treatment via $I_{Kr}$ block.}} {\bf{a)}} Traces of the rapid component of the delayed rectifier K$^{+}$ current for the models calibrated to the SR dataset (blue, solid) and the cAF dataset (red, dashed). The cAF population demonstrates an almost twofold increase in the activity of this current. {\bf{b)}} APs after treatment by 50\% $I_{Kr}$ block (gold) show significant AP prolongation compared to the same models without $I_{Kr}$ block (red). This is also demonstrated by the averaged traces for both (black lines, treated -- solid, untreated cAF -- dashed). The restoration of atrial refractoriness in patients with cAF is clearly demonstrated.}
\label{fig:iKr}
\end{figure}

 In our POMs calibrated to the SR and cAF datasets, the cAF models show significantly larger $I_{Kr}$ that also activates slightly earlier (\figref{fig:iKr}a), contributing to the more rapid repolarisation and lack of a plateau phase in the cAF APs. However, the maximum conductance of the channel, $g_{Kr}$, shows no significant difference between the SR and cAF POMs (see \tabref{tab:ATR}). This indicates that the increase in $I_{Kr}$ activity is symptomatic of the changed AP morphology in cAF, which impacts the voltage-dependent gating behaviour of these ion channels.

We explore $I_{Kr}$ block via drug treatment by first pacing the models in the cAF POM until steady state (see Materials and Methods), then reducing $g_{Kr}$ by 50\% and repeating the full stimulus protocol. In correspondence with the observed effects of such drug treatments, the APDs of almost all models (95\%) are restored to values associated with healthy SR (\figref{fig:iKr}b). This demonstrates the ability of our calibration to AP biomarker measurements to produce models that exhibit appropriate behaviours even in situations outside of those to which they were calibrated. However, there are a small number of models that predict \textit{decreased} APD in response to treatment, as shown by the post-drugblock APs (gold) that fall to the left of the untreated cAF APs (red), as well as a few models that repolarise to unreaslitically high resting potentials ($>\!-60\:\:\!\!\mbox{mV}$). One advantage of the POMs framework is that these models that show unexpected behaviour can be directly examined in order to determine the underlying causes, potentially identifying risk factors for these adverse reactions.

The models that repolarise extremely rapidly following treatment with $I_{Kr}$ blocker are seen to be associated with very small values of $g_{NaK}$, and thus the unexpected behaviour of these models likely stems from the selection of values for this parameter which are too small to be physiologically realistic. This is a risk of choosing such a large ($\pm 100\%$) extent of variation of our parameter values, further motivating our exploration of a lower extent of variability in the following section. Nevertheless, these models do identify a potential risk factor for the use of $I_{Kr}$-blocking treatments, namely that insufficient $I_{NaK}$ activity can result in dangerous further reduction of the refractory period. Examining the current activity in these models reveals significant Na$^{+}$ ion accumulation that occurs due to the reduced action of $I_{NaK}$, which is further hampered by the reduced flow of K$^{+}$ ions through $I_{Kr}$. This then triggers extreme currents outward through the Na$^{+}$/Ca$^{2+}$ exchanger, resulting in the extra-rapid repolarisation that is observed.

The models repolarising to resting potentials that are unrealistically high are seen to be associated with incomplete deactivation of the L-type calcium channels and lower values of $g_{K1}$, resulting in an imbalance of inward and outward current in the unexcited cell that gradually pushes up its membrane potential at rest. Eventually, an alternative steady state is reached where the elevated resting potential largely prevents the cell's sodium channels from opening and the AP is severely disrupted. Parameter values that lead to this behaviour will never be selected by the calibration process, because their APA and RMP values fall outside of the data. However, when outward current due to $I_{Kr}$ is reduced by drug treatment and the balance of ion flow is changed, a few of the models then fail to achieve correct homeostatic balance and instead end up at this alternative steady state. The questions of whether other AP models also predict such alternative steady states, and whether drug treatments can cause individual atrial cells to develop disrupted balances of ion flow at rest (compensated for by their neighbours), are beyond the scope of this paper.

The drugblock case study can thus be seen as a means of further calibrating the generated POMs, by testing the ability of all of the models selected to continue predicting reasonable AP curves when subject to established treatments. This is important given the tendency for currents to compensate for one another, resulting in model behaviours that only become manifest subject to this type of further interrogation. In the previous study of Britton \textit{et al.} \cite{Britton2013}, albeit using a different AP model, range-based calibration to biomarkers recorded for different pacing frequencies was seen to be sufficient for avoiding the selection of models that exhibit unphysical responses to drugblock.

\subsubsection*{(iii) Calibration of Model Parameters in Response to Variable Data}

Although calibrated POMs are particularly suited to explaining variable data by creating representative populations, the technique can also be used to select a single set of parameter values for a given model in response to available data. This is particularly appropriate when variability in data is expected to source from uncertainty introduced by the experimental process, or by other means not explained by model parameters. Problems of this nature are \textit{inverse problems}, and ubiquitous in a wide range of fields \cite{Aster2005}. Although numerous solution techniques for such problems exist, an advantage of calibrated POMs in this context is that the space of parameter values that generate outputs close to the data is provided as an output, similar to Bayesian approaches that generate a posterior distribution for the parameters \cite{Aster2005, Johnstone2016}. Taking into account the range of different parameter values that generate outputs close to the data, as opposed to simply finding a set of parameters that generates an optimal output is important both for supplying uncertainty estimates on parameter values and dealing with models where disparate parameters can generate very similar outputs.

POMs calibrated successfully to data density represent a set of models that all correspond in a sense to some portion of the data. Therefore, generating a single set of parameter values from a distribution-calibrated POM is simply a matter of selecting an appropriate means of condensing this set of models back into a single set of parameter values. In our case, the set of parameter values selected for both the SR and cAF POMs have a rather regular distribution (no obvious bimodal behaviour or obvious correlation structures, see \figref{fig:theta_distributions}), and so we take the median values of each individual parameter to create two modified CRN models, one each for the SR and cAF datasets.

\begin{figure}
\begin{center}
\includegraphics[width=14cm, trim={4cm 0.5cm 4cm 1cm},clip]{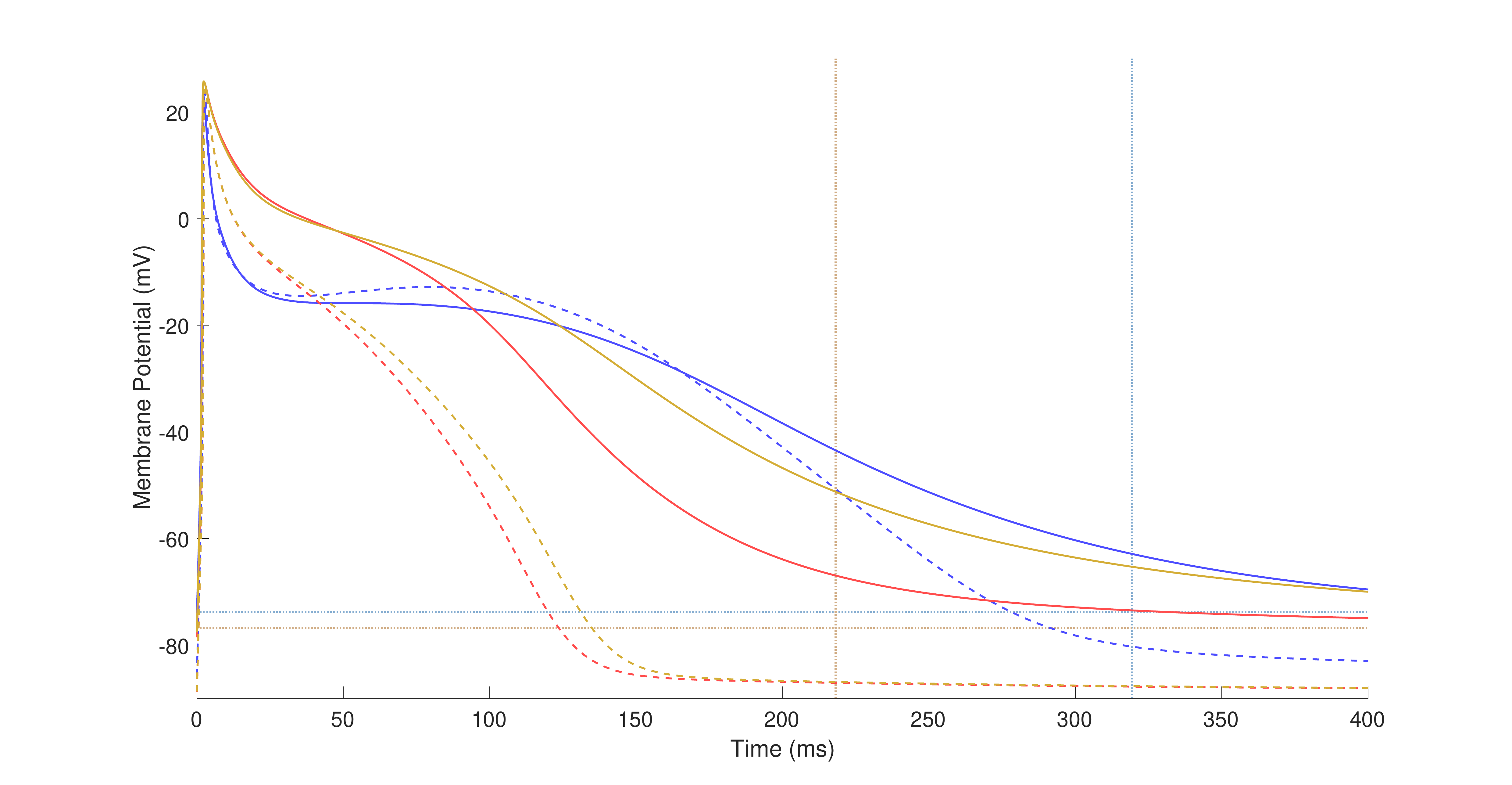}
\end{center}
\caption{{\bf Selection of parameter values using distribution-calibrated POMs produces updated models that correspond to provided data.} SR (blue) and cAF (red) APs, and the effects of 50\% $I_{Kr}$ block (gold) as predicted by the base CRN model (dashed lines) and experimentally-calibrated CRN models using the median of POM $\boldsymbol{\theta}$ values (solid lines). The former predict significantly lower RMP and APD$_{90}$ than suggested by the data (mean values from the data for these biomarkers are indicated by guidelines, SR -- blue, cAF -- red). The modified CRN models correctly predict RMP and APD$_{90}$ values that correspond to the data, and also successfully capture the antiarrhythmic effect of $I_{Kr}$ block, demonstrating the restoration of an SR APD in the cAF model.}
\label{fig:singlemodels}
\end{figure}

Figure \ref{fig:singlemodels} compares our modified CRN models to the baseline CRN model, and the baseline CRN model modified for cAF using experimentally-informed adjustments to its parameters taken from \tabref{tab:ATR}. It can immediately be seen that the original CRN model for both SR and cAF underestimates the RMP and APD$_{90}$, and that our modified CRN models rectify this issue very well. More notably, the responses to drugblock treatment predicted by the original and modified cAF models are completely different, with our calibrated model demonstrating the correct restoration of APD to SR in response to $I_{Kr}$ block while the original CRN model demonstrates only very minor APD prolongation. The original CRN model's lack of response to $I_{Kr}$ block is a result of the model's prediction of reduced $I_{Kr}$ activity in cAF \cite{Courtemanche1999}, in contradiction to the increase predicted by our POMs (\figref{fig:iKr}a). Given the known efficacy of this type of treatment for the restoration of SR in patients with cAF, we suggest that our technique of updating parameter values in response to new experimental data using distribution-calibrated POMs can also produce more predictive models in the case of for example drug treatments.

\subsection*{Distributional Calibration Can Inform the Extent of Variability in Parameter Values}

Calibrating to distributions means that the variability in a supplied dataset can be expected to be explicitly captured by constructed POMs, so long as the model across the specified parameter space is capable of generating outputs that match the data. This means that by using divergence measures such as $\rho$ and $\hat{\rho}$, the differential ability of various parameter spaces to capture variable data with POMs can be considered. The most obvious application of this technique is to explore the level of variability in parameter values needed to explain the data.

Our studies on the Sanchez \textit{et al.} dataset have used a large variation in ion channel conductance values ($\pm 100\%$), following previous studies \cite{Britton2013, Sanchez2014, Zhou2013}. Working with such a variation in parameter values carries the risk of selecting extreme values that are not physiologically sound (for example values close to $-100\%$ that essentially switch off an entire current), and there is some suggestion that a level of variability such as $\pm 30\%$ is more appropriate for cardiac ion channel conductances \cite{Romero2009, Sanchez2014, Muszkiewicz2016}. We therefore seek to answer the question of whether 30\% variability sufficiently explains the variation in the dataset we calibrated to.

\begin{figure}
\begin{center}
\includegraphics[width=15cm]{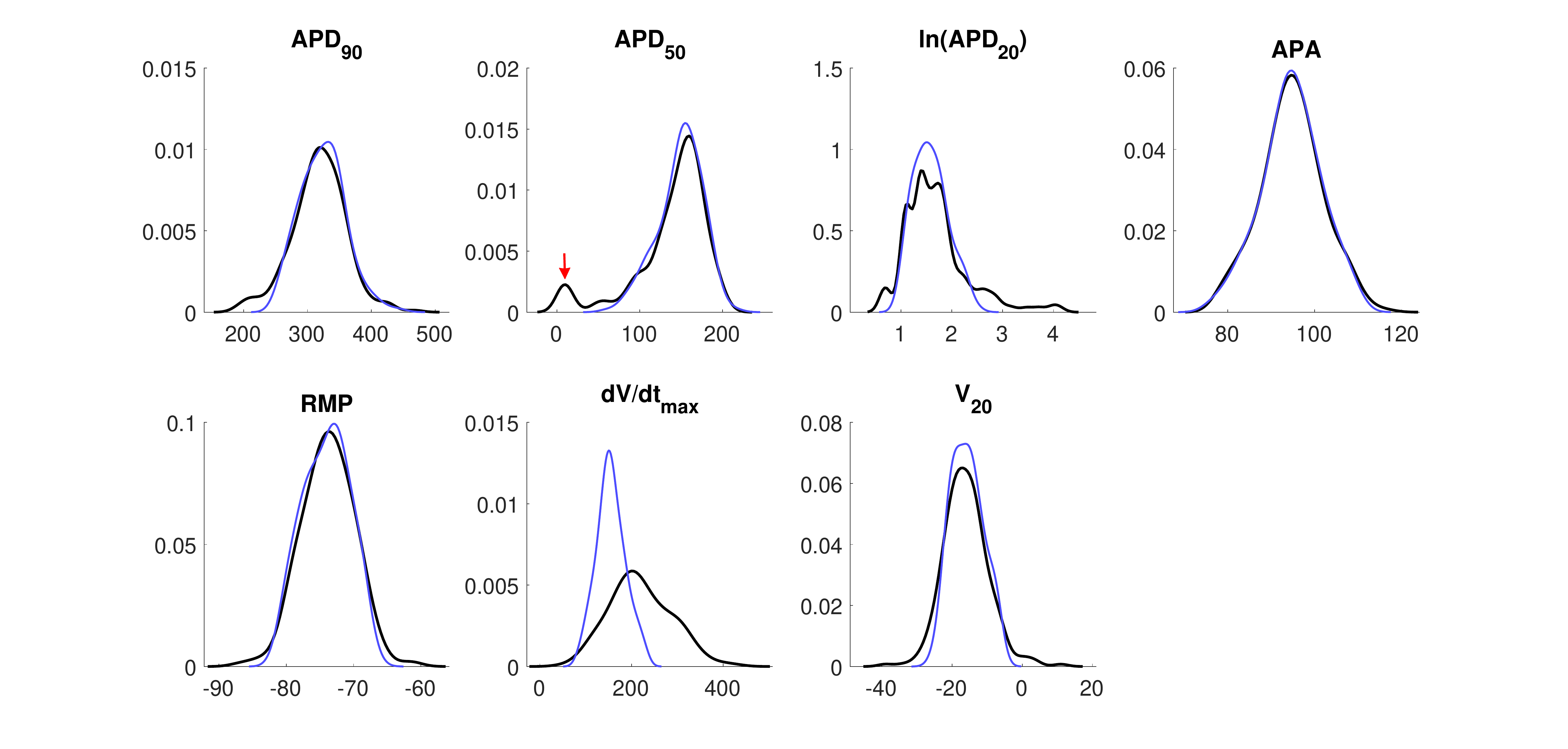}
\end{center}
\caption{{\bf Variation of $\mathbf{\pm 30\%}$ in current densities underestimates biomarker variance in the SR dataset.} Marginal distributions of the biomarkers in the SR dataset (black) and distribution-calibrated POM using $\pm 30\%$ variance in ion channel conductances (blue). A reduced search space is still able to recover the general distributions of all biomarkers except for $\de{V}{t}_{\mbox{\scriptsize max}}$, but the full extent of variation in the APD biomarkers and V$_{20}$ is not present in the calibrated POM. Notably, the very low APD$_{50}$ values recorded for some patients are completely unrepresented in the POM (arrow).}
\label{fig:SR_30pc_marginals}
\end{figure}

\begin{figure}[htp]
\begin{center}
\includegraphics[width=15cm, trim={4cm 2cm 4cm 1cm}, clip]{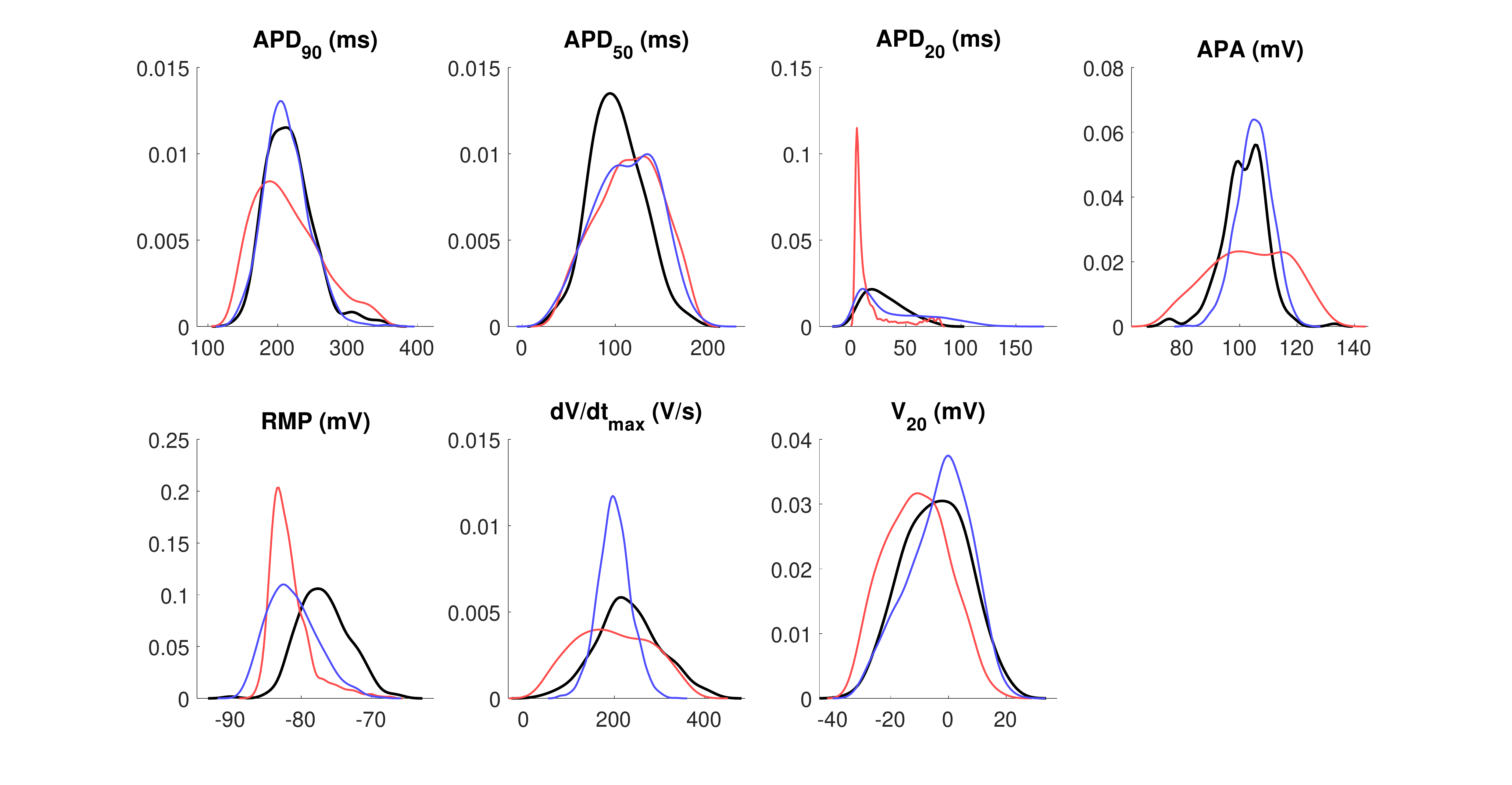}
\end{center}
\caption{{\bf{Calibration to biomarker distributions, as opposed to their ranges, significantly reduces model bias for the cAF dataset.}} Marginal distributions of the biomarkers in the cAF dataset (black) and POMs calibrated to biomarker distributions using the SMC algorithm (blue) or calibrated to biomarker ranges using LHS (red). Calibration to biomarker distributions, as opposed to their ranges, significantly reduces  model bias and produces a much more representative POM. This is  demonstrated by the marginal distributions of POMs constructed with LHS matched to ranges (red) and SMC calibrated to the biomarker distribution (blue) in comparison with the marginal distributions of the biomarkers in the cAF dataset (black). SMC demonstrates a significant improvement in capturing the distributions of almost all biomarkers.}
\label{fig:SMCvsLHS_CA_marginals}
\end{figure}

\begin{figure}
\begin{center}
\includegraphics[width=15cm, trim={4cm 2cm 4cm 1cm}, clip]{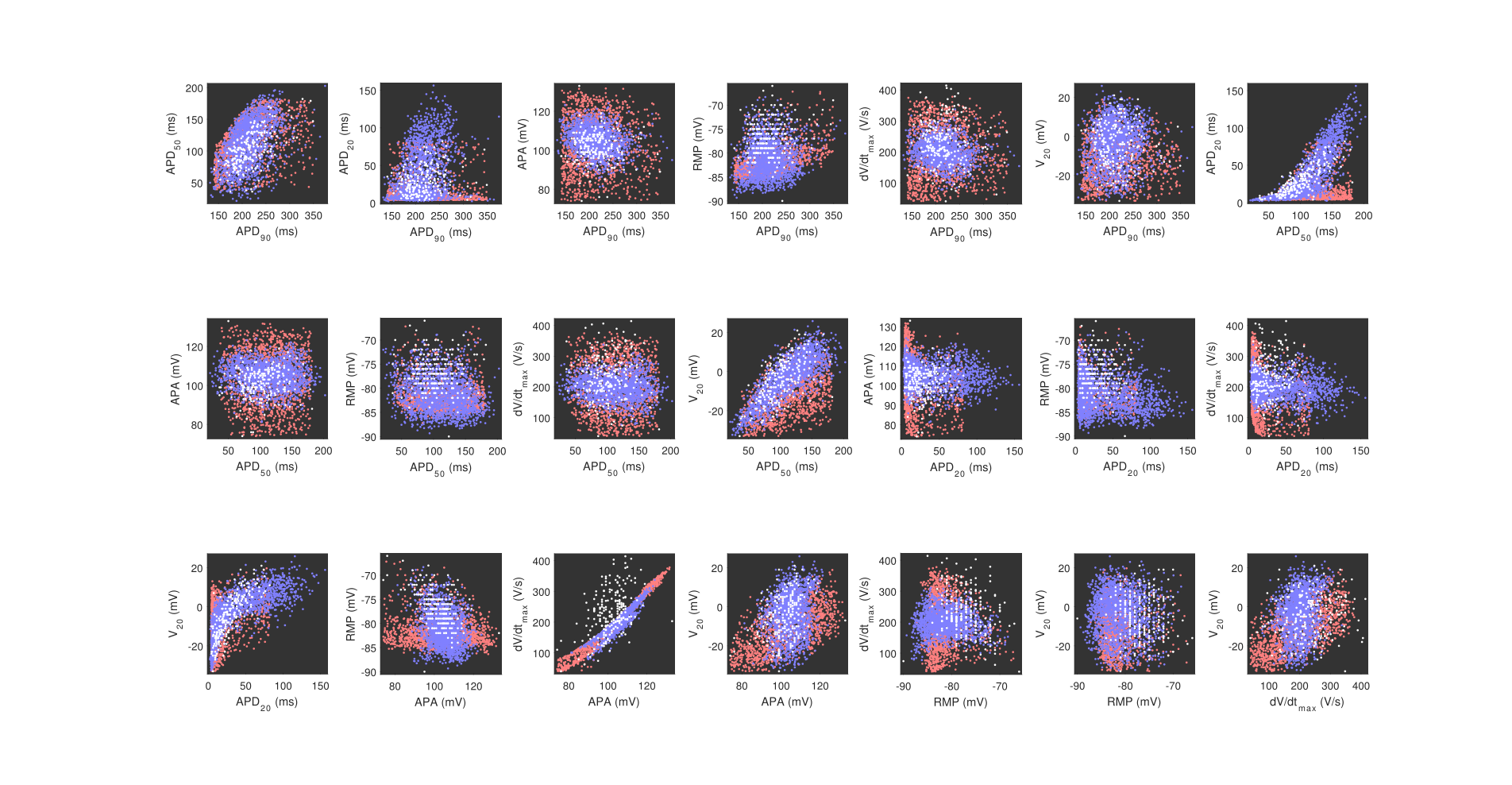}
\end{center}
\caption{ {\bf{Bivariate distributions of biomarker pairs are well captured by an SMC-constructed POM for the cAF dataset.}} Pairwise scatterplots of each unique pair of biomarkers in the SR dataset (white) and the POMs constructed using SMC matched to distributions (blue) and LHS matched to ranges (red). The SMC-generated POM demonstrates better localisation to the dense regions in the data, but clearly requires further calibration. The same very strong correlation between APA and $\de{V}{t}_{\mbox{\scriptsize max}}$ seen for models selected to match the SR dataset is also seen here.}
\label{fig:SMCvsLHS_CA_biopairs}
\end{figure}

\begin{figure}
\begin{center}
\includegraphics[width=15cm, trim={4cm 2cm 4cm 1cm},clip]{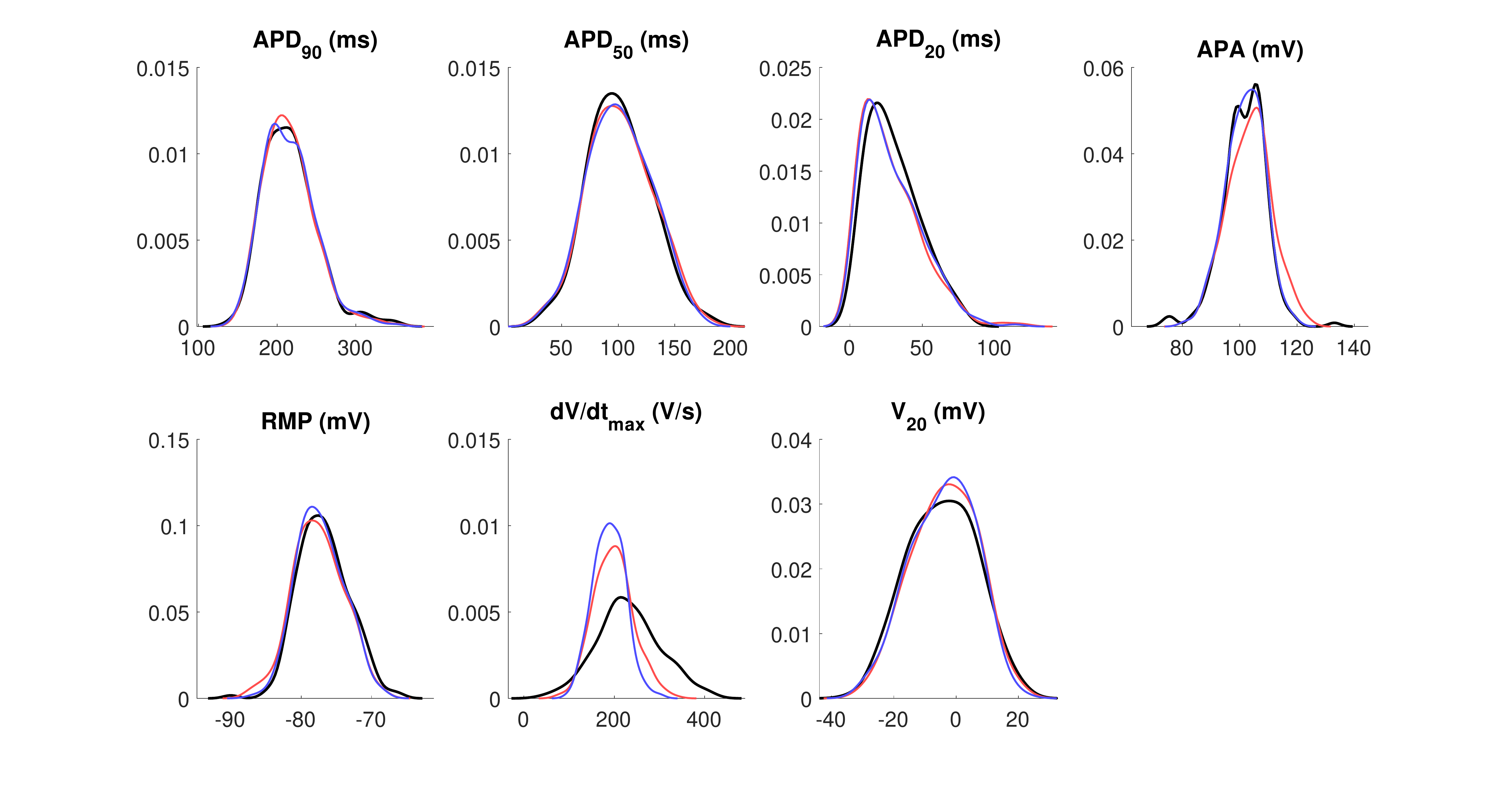}
\end{center}
\caption{{\bf{Variability in the cAF dataset is captured by a population of CRN models with varying current densities.}} Marginal distributions of the biomarkers in the cAF dataset (black) and the POMs constructed using SMC followed by simulated annealing to minimise $\rho$ (red) or $\hat{\rho}$ (blue). Matching of the univariate biomarker distributions is slightly less well achieved than in the case of the SR datsaet, but the calibration process is clearly very successful and the trends in the data captured by the constructed POM.}
\label{fig:CA_marginals}
\end{figure}

\begin{figure}
\begin{center}
\includegraphics[width=15cm, trim={4cm 2cm 4cm 1cm},clip]{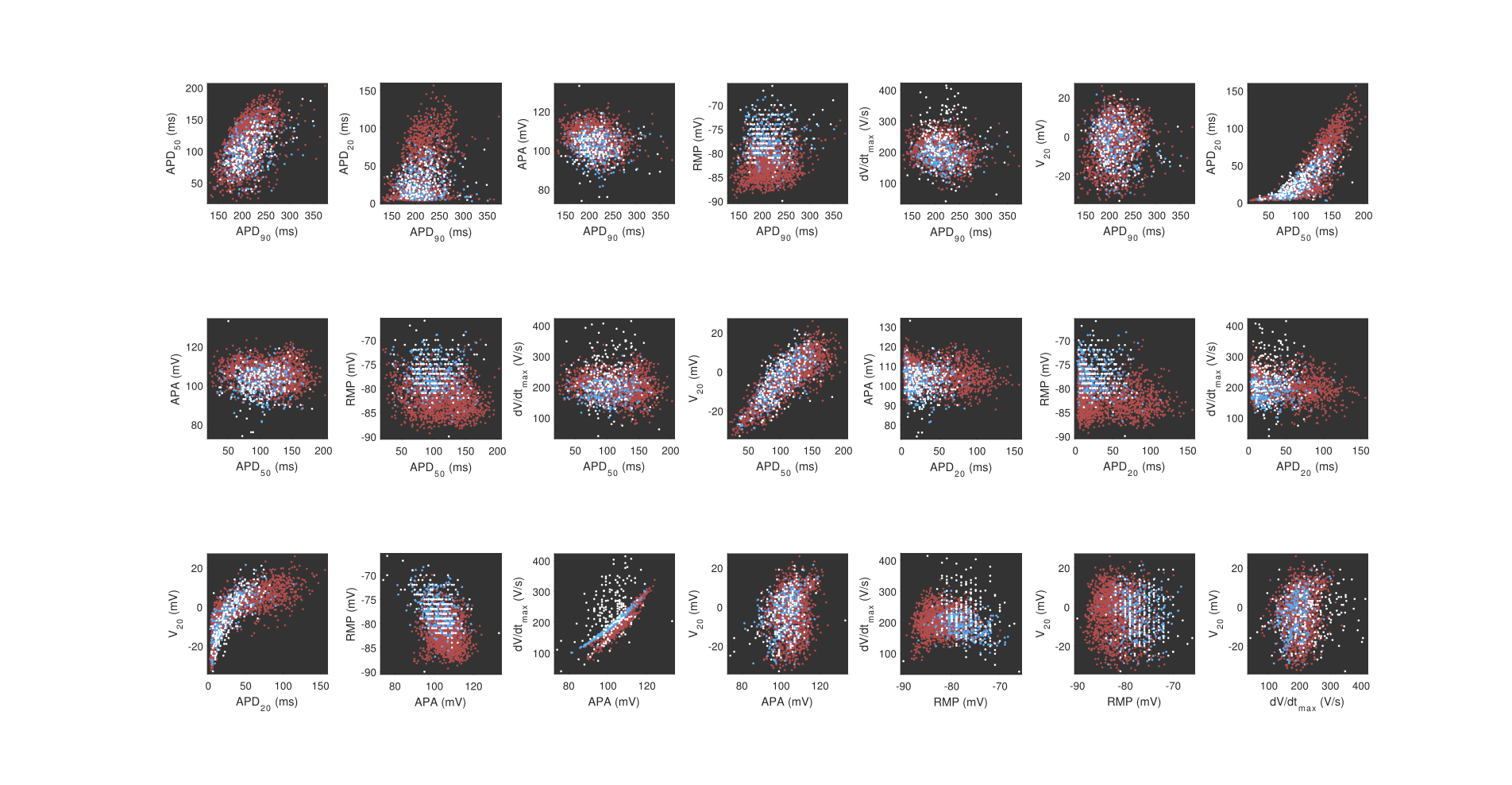}
\end{center}
\caption{{\bf{Simulated annealing successfully selects models according to data density in the biomarker space in the cAF dataset.}} Pairwise scatterplots of each unique pair of biomarkers in the cAF dataset (white) and the models from the SMC-generated POM that are accepted (light blue) or rejected (grey) in the process of minimising $\hat{\rho}$. Only the spread of, and correlations with, $\de{V}{t}_{\mbox{\scriptsize max}}$ are not captured very well by the final POM.}
\label{fig:CA_biopairs}
\end{figure}

\begin{figure}
\begin{center}
\includegraphics[width=12cm, trim={4cm 0.5cm 4cm 1cm},clip]{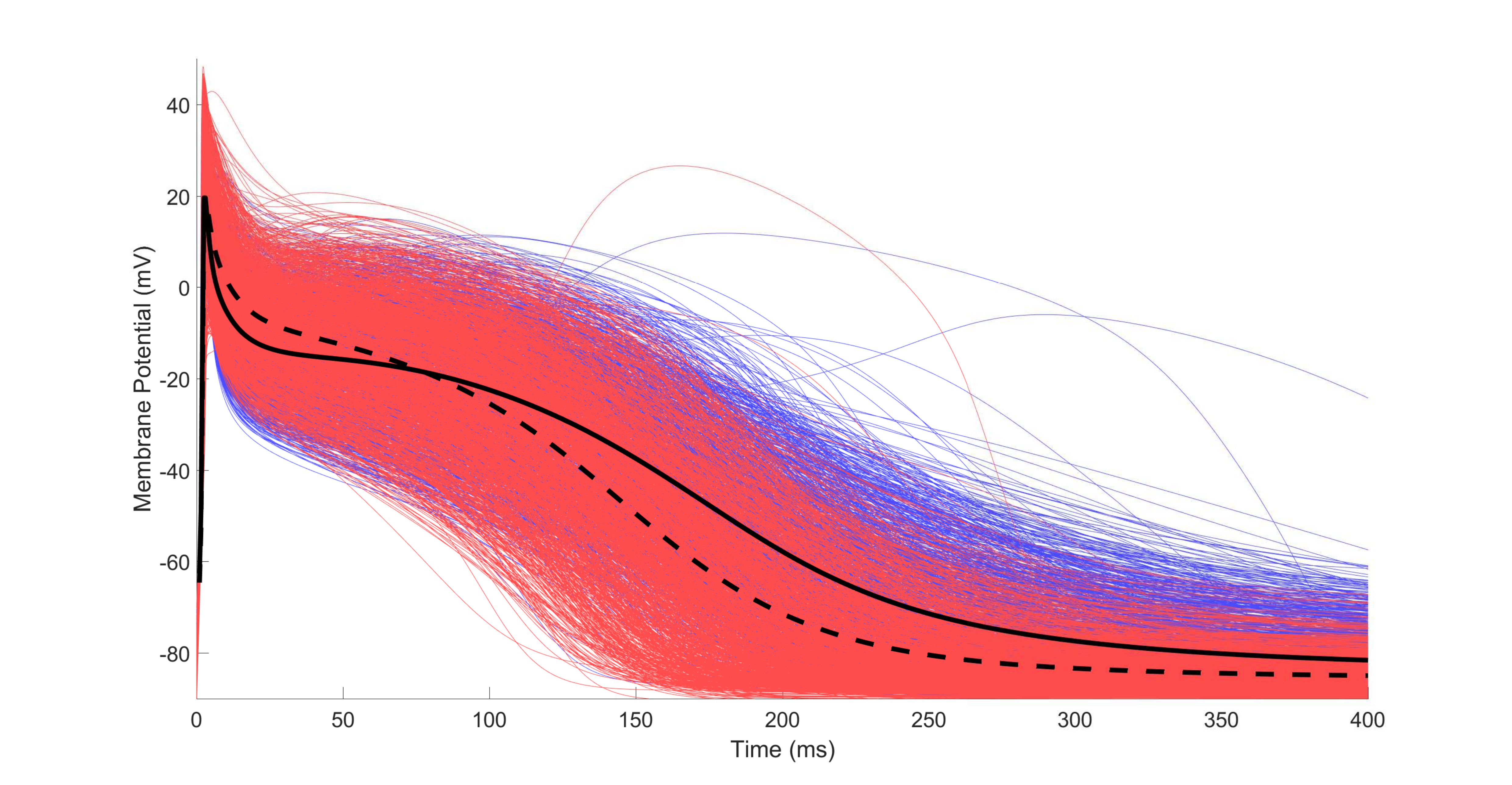}
\end{center}
\caption{{\bf{Calibration to ranges fails to capture the morphological differences between SR and cAF atrial action potentials.}} Atrial action potentials produced by simulation of the populations of CRN models calibrated to the ranges of biomarker data for patients exhibiting sinus rhythm (blue) and chronic atrial fibrillation (red). Also displayed are the average of all traces for the sinus rhythm (solid) and atrial fibrillation (dashed) populations. Differences in AP morphology are far less pronounced than those observed using calibration to distributions, and a small number of simulated APs appear unphysical.}
\label{fig:AP_traces_LHS}
\end{figure}

\begin{figure}
\begin{center}
\includegraphics[width=15cm, trim={6cm 2cm 4.5cm 1cm}, clip]{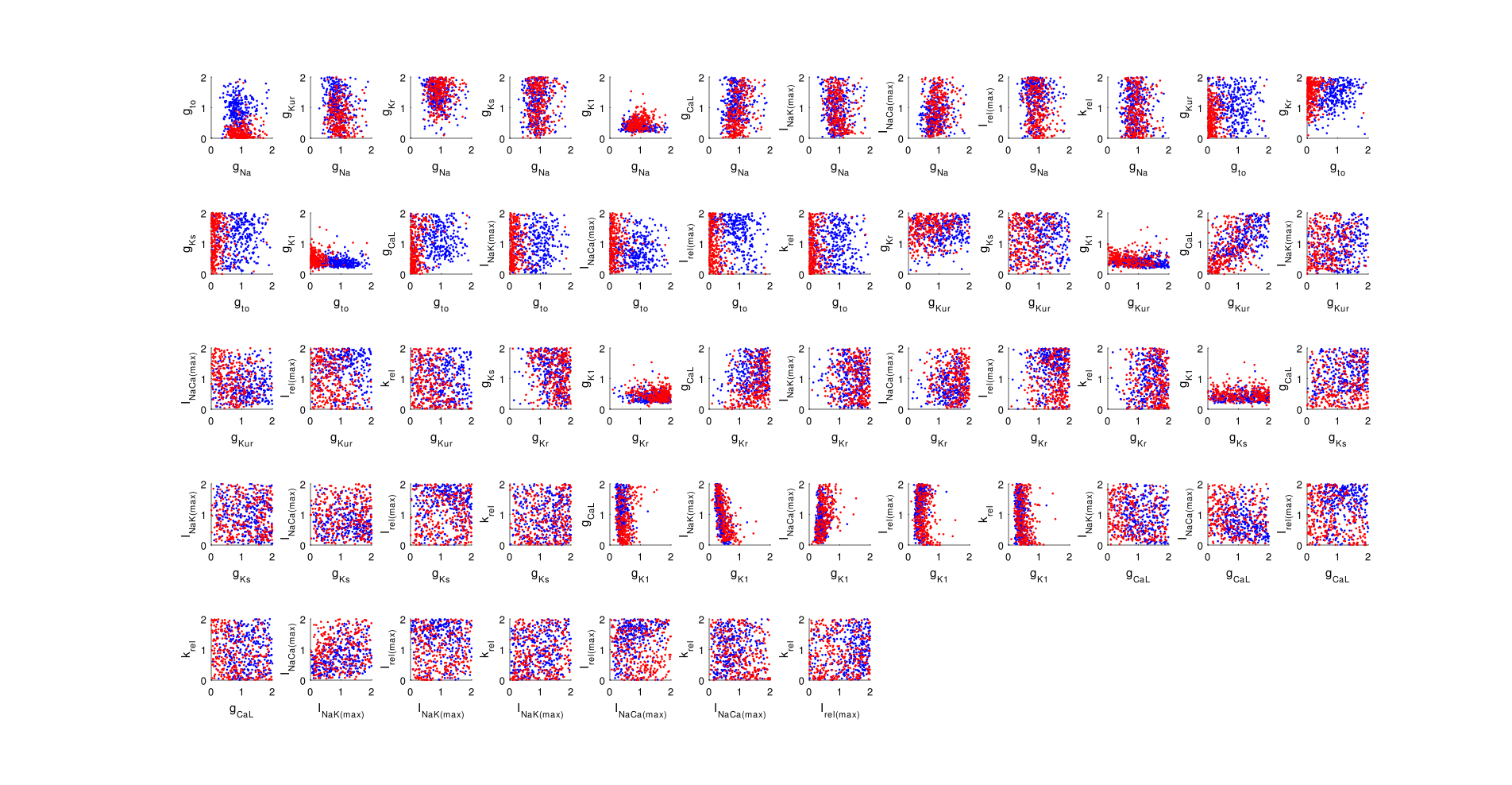}
\end{center}
\caption{{\bf The distributions of parameter values selected for the SR and cAF POMs are distinct, but regular.} Pairwise scatterplots of the parameter values selected for POMs calibrated to the SR (blue) and cAF (red) datasets, expressed in terms of the proportion of the base values for parameters in the CRN model. Clear differences in the two distributions can be observed, but neither POM exhibits obvious patterns of correlation in any pair of parameters, nor is there evidence of bimodality. These properties are important when reducing a POM back to a single representative model.}
\label{fig:theta_distributions}
\end{figure}

\begin{figure}
\begin{center}
\includegraphics[width=15cm, trim={4cm 0.5cm 4cm 1cm}, clip]{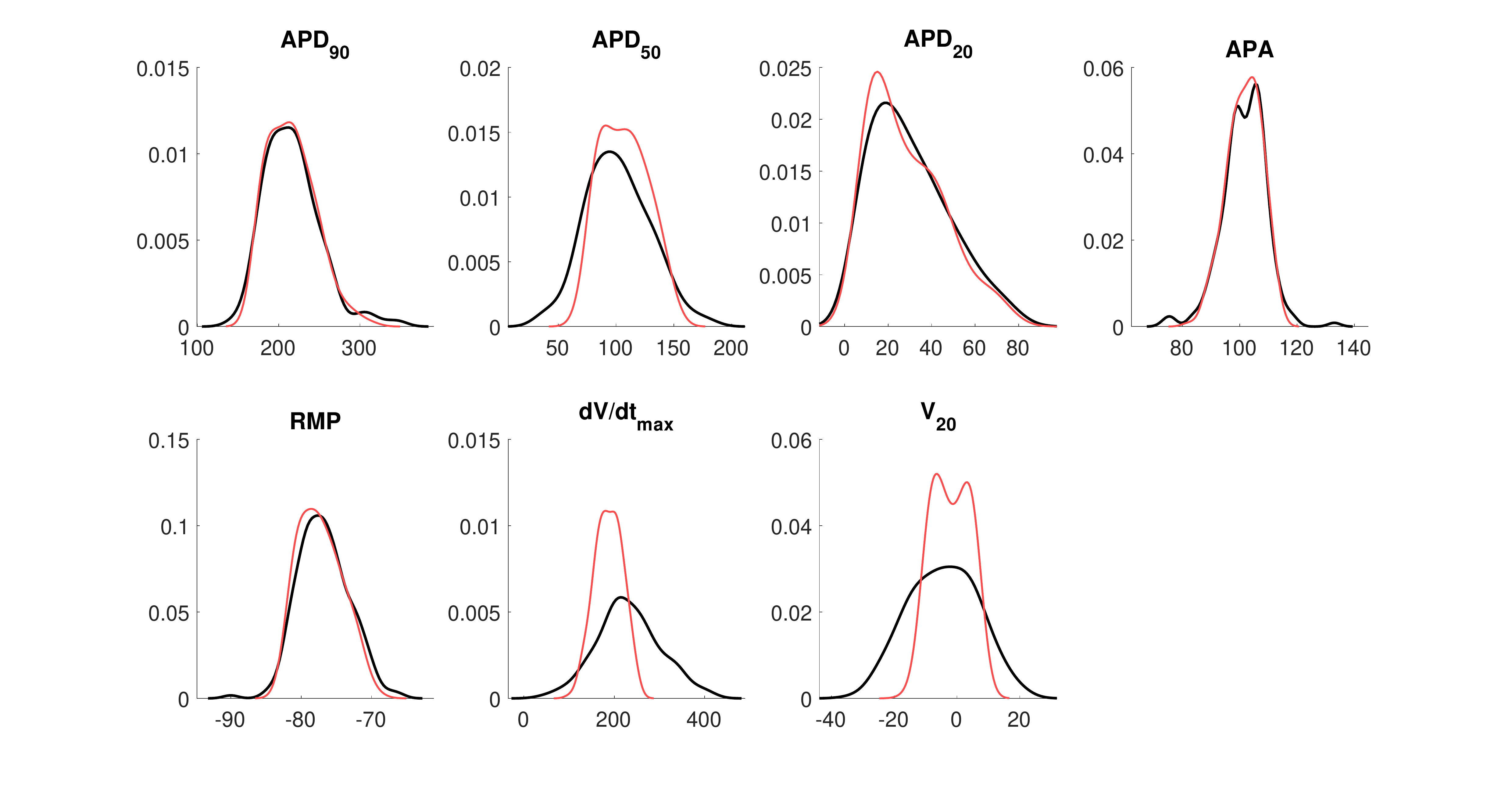}
\end{center}
\caption{{\bf Variation of $\mathbf{\pm 30\%}$ in current densities underestimates biomarker variance in the cAF dataset.} Marginal distributions of the biomarkers in the cAF dataset (black) and distribution-calibrated POM using $\pm 30\%$ variance in ion channel conductances (red). A reduced search space is still able to recover the general distributions of all biomarkers except for $\de{V}{t}_{\mbox{\scriptsize max}}$ and V$_{20}$, with the extent of variation in APD$_{50}$ also significantly underestimated.}
\label{fig:CA_30pc_marginals}
\end{figure}

In order to select the most relevant portion of the parameter space, we take parameter values $\pm 30\%$ around the values selected for the modified CRN models in the previous section. Applying our SMC sampling algorithm and subsequent refinement,  
we obtained 254 models calibrated to the SR dataset, and 215 models calibrated to the cAF dataset, the minimum allowable number of models in both cases. Neither of these POMs succeeded in fully capturing the variability in the dataset, resulting in divergence measures that compare unfavourably to those obtained using the full $\pm 100\%$ variation in cell properties (\tabref{tab:rho30pc}). Examination of the marginal distributions of the biomarkers for the SR $\pm 30\%$ POM reveals that it does successfully capture the general distributions of the data, but fails to show the same extent of variance (\figref{fig:SR_30pc_marginals}). Similar results are also seen for the cAF $\pm 30\%$ POM (\figref{fig:CA_30pc_marginals}).

\begin{table}[ht]
\begin{center}
\begin{tabular}{lcccccccc}
\hline
\bf{Population of Models} & & \multicolumn{3}{c}{\bf{SR data}} & & \multicolumn{3}{c}{\bf{cAF data}}\\
 & & $\rho$ & & $\hat{\rho}$ & & $\rho$ & & $\hat{\rho}$ \\ \hline
$\pm 100\%$ parameter space $\hat{\rho}$ & & 1.47 & & 0.49 & & 1.28 & & 0.57 \\
$\pm 30\%$ parameter space $\hat{\rho}$ & & 1.79 & & 1.01 & & 1.77 & & 1.19 \\
\hline
\end{tabular}
\end{center}
\caption{{\bf{Variability of $30\%$ in ion channel conductances fails to fully explain the variable data.}} Comparison of the ability of different POMS to capture the between-subject variability in two clinical datasets, when the extent of underlying variability in ion channel conductance is decreased. The larger $\rho$ and $\hat{\rho}$ values seen for the $\pm 30\%$ POMs indicate a more significant divergence between POM and data, resulting from the inability to find models in this smaller parameter space that produce all of the combinations of biomarker values in the dataset.}
\label{tab:rho30pc}
\end{table}

Furthermore, the secondary mode in the distribution of SR APD$_{50}$ values is completely unrepresented by the POM constructed with reduced variability. This peak most likely corresponds to cells that repolarise more than 50\% during the phase of immediate repolarisation, driven primarily by the rapidly activated outward currents $I_{to}$ and $I_{Kur}$. This results in a cluster of very small APD$_{50}$ values that are separate from the majority of cells, which only reach 50\% repolarisation after the plateau phase. General underestimation of biomarker variability in the POM could potentially be explained by measurement error associated with the biomarker data, but the inability to produce models that populate this peak implies that additional variance in cell properties, at least those relevant to $I_{to}$ and/or $I_{Kur}$, are required to produce models that exhibit sufficient early repolarisation. Coupled with the general underestimation of biomarker variance in the POM, these results certainly imply that additional variability in ion channel conductances beyond $\pm30\%$ underlies the variability in biomarkers that is seen in these datasets.

\section*{Discussion}

Population-based modelling is a powerful technique for allowing deterministic mathematical models to explore and characterise the variability inherent in complex systems \cite{Sarkar2012}. This includes the use of multiple regression techniques performed on synthetic populations \cite{Sobie2009, Sarkar2010, Sarkar2011}, and when data is explicitly available, calibrated POMs \cite{Muszkiewicz2016}. Previously, calibration of POMs has been performed by ensuring that all relevant model outputs fall within the ranges of the data \cite{Britton2013, Sanchez2014, Gemmell2016}. While this is a perfectly reasonable approach, especially when a low number of experimental samples is available, it does ignore other information inherent in the data and does not strictly guarantee that the models selected will produce outputs that correspond to those seen in the data. Our presented calibration technique estimates the underlying distribution of outputs represented by the data and calibrates to this distribution using a combination of SMC \cite{Drovandi2016} and a simulated annealing-type algorithm.

We have demonstrated the efficacy of our technique on a pair of datasets from cardiac electrophysiology, a field that has seen a great deal of POMs research following Prinz, Bucher and Marder's \cite{Prinz2004} and Marder and Taylor's \cite{Marder2011} pioneering studies in neuroscience. The data used was comprised of biomarker values collected for atrial myocytes from hearts exhibiting healthy SR or the cAF pathology \cite{Sanchez2014}, offering two sets of data to test the calibration technique and the opportunity to compare the models selected to represent healthy and afflicted hearts. Use of SMC to sample according to the distribution of data was found to improve the correspondence between POMs and experimental observations over LHS matched to ranges, but the approximate nature of the technique left room for further improvement. A subsequent refinement technique to select optimal subpopulations of the SMC-constructed POMs was able to fully recover the variability in the dataset, except where the underlying model was incapable of producing the outputs that were experimentally observed.

Our approach generates models that predict very well the morphologies of different types of APs associated with SR or cAF. The cAF models also demonstrate a more realistic response to antiarrhythmic treatment (AP prolongation in response to $I_{Kr}$ block \cite{Colatsky1994, Savelieva2008}), as compared to the original CRN model. These benefits remained even when the population of models was averaged to produce a single model, demonstrating the efficacy of the technique for the more general problem of selecting parameter values in response to variable data. Taking into account the variability in data in this fashion, as opposed to simply averaging it and fitting parameters to the result, is of particular importance when a system (and its associated mathematical model) is complex, such as in our example of cardiac electrophysiology where the ability of currents to compensate for one another allows for highly similar APs to be produced by very different balances of constituent currents \cite{Britton2013, Mann2016}.

Furthermore, calibration that takes more of the features of data into account may be able to resolve more subtle differences between datasets, making it a powerful approach for identifying the potential causative factors that produce these differences. In our case, calibrating POMs to the distribution in the SR and cAF datasets was able to select POMs with very distinct parameter values, and these differences in parameter values were seen to be indicative of the known changes in atrial myocyte behaviour associated with the cAF pathology \cite{Dobrev2003, Pandit2013}. Interestingly, differences in the action of the sarcoplasmic reticulum were also identified, despite the uptake and release of Ca$^{2+}$ internally not contributing directly to the AP \cite{Courtemanche1998}. We suspect that the contribution of the internal calcium dynamics to the current through the Na$^{+}$/Ca$^{2+}$ exchanger allows them to be partially identified, even when calibrating only to AP biomarkers, but this point requires further investigation beyond the scope of this paper. 

In this case, following the available data, we have used the well-studied cAF pathology to demonstrate our density-based calibration technique's ability to detect the differences in parameter values that underlie different datasets. However, appropriately-calibrated POMs could also be used to potentially identify differences in behaviour (codified by parameter values) that explain pathological conditions that are not as well understood. This same thought also extends to improved identification of any differences that underlie artificial stratifications of a dataset, for example differences in physiology that correspond to factors such as gender or age \cite{Taneja2001, Sakabe2003}.

When constructing POMs without experimental evidence for the extent of variability in parameters, the choice of parameter space to sample is open. Our primary study used a large space of parameter values ($\pm 100\%$) to give the model the best opportunity to simultaneously fit the two different sets of data and hence demonstrate our technique. However, we also considered the effects of choosing a smaller parameter space ($\pm 30\%$, \cite{Romero2009, Walmsley2013, Sanchez2014}), attaining results that imply that $\pm 30\%$ variance in primary current conductances is insufficient to explain the experimental data, and thus that further variation in these cell properties or others is expected. This sort of exploration into appropriate levels of variability is particularly important in fields such as cardiac electrophysiology, where the variable cell properties are difficult to directly measure and the extent of variability is not well established \cite{Mirams2016}. Such analysis requires the ability to calibrate POMs to distributions, so that the comparisons between parameter spaces of different sizes are unaffected by model bias.

There are several circumstances under which our calibration process might fail or be considered inappropriate. Firstly, calibrating to distributions requires sufficient data in the dataset to form a reasonable approximation to the underlying distribution. When the sample size is insufficient compared to the variance to suggest a specific distribution with any real confidence, obviously distributional calibration is inappropriate and calibration based on range statistics is very reasonable. However, even in these cases a benefit can potentially be gained by enforcing that data is spread evenly across its range, although we have not investigated this in this paper.  Secondly, when the model itself is incapable of producing outputs that correspond to some experimental data, no calibration process will allow it to capture these portions of the data. In this scenario, the model itself must be examined, or the experimental protocols  reconsidered. This does however demonstrate the usefulness in considering the distribution of the data and how well constructed POMs fit to it using measures such as $\rho$, in that it suggests when models (or possibly experiments) might need to be reconsidered. Indeed the varying ability of different models to successfully produce POMs calibrated to a given dataset also serves as a means of comparing and benchmarking them. Lastly, the reliance upon a transformed target density in the SMC algorithm might potentially cause it to select models that do not accurately capture the distribution of outputs in the data, even though on our test problem the method has been seen to perform very well. In the worst case, `naive' POMs composed of large numbers of models can be constructed by sampling the search space uniformly (such as through LHS) and then refined by our simulated annealing process to select the subpopulation of these that best matches the data. 

In conclusion, our presented calibration technique allows datasets to be thoroughly mined in order to produce informative and predictive POMs that capture the variability between individual members of a population. The benefits of successfully accounting for this variability are well established \cite{Sarkar2012}, including the selection of parameter values in response to variable data, uncovering differences in underlying behaviour that characterise different datasets (or stratifications within a single dataset), identification of factors that may predict the differential response to drug treatments, and better informed comparisons between different models for the phenomena in question. We have demonstrated how POMs calibrated to the distributions in data using our approach can successfully perform these functions using a real dataset in atrial electrophysiology. The technique is best suited to datasets that are large enough to clearly suggest an underlying distribution of the quantities being observed, but calibration to smaller datasets could potentially be achieved by fitting to uninformed distributions that simply ensure that the resultant outputs of constructed POMs are sufficiently spread across the range of potential values.

\section*{Materials and Methods}
\subsection*{Experimental Dataset}
\label{sec:methods_data}

The clinical data used in this work was that presented by Sanchez \textit{et al.} \cite{Sanchez2014}, namely biomarker values measured from recorded action potentials for 469 cells taken from the right atrial appendages of 363 patients. These biomarker readings were split into two groups, patients exhibiting standard sinus rhythm (SR) and patients exhibiting chronic atrial fibrillation (cAF). The biomarker values used to quantify the action potentials were the action potential durations compuated at 20, 50 and 90\% repolarisation (APD$_{20}$, APD$_{50}$, APD$_{90}$), the action potential amplitude (APA), resting membrane potential (RMP), the potential at 20\% of APD$_{90}$ (V$_{20}$) and the maximum upstroke velocity (dV/dt$_{\mbox{\scriptsize max}}$). More information regarding the experimental conditions under which the data was collected is available in \cite{Sanchez2014}.

The data demonstrates statistically significant differences between biomarker values for SR and cAF cells, with $p < 0.001$ for all biomarkers except dV/dt$_{\mbox{\scriptsize max}}$ \cite{Sanchez2014}. The relatively large size of the dataset allows for the specific distributions of each biomarker, and their dependencies on each other, to be meaningfully explored.

We also note that this dataset was ideal for the development of our new POM calibration technique based on capturing the distributional features of a dataset in that biomarker readings were collected from over 200 myocytes from patients in both SR and cAF and therefore sufficient data is available to make distributional calibration both meaningful and appropriate.

\subsection*{Atrial AP Model}
\label{sec:methods_model}
This work used the Courtemanche--Ramirez--Nattel (CRN) model for atrial action potentials \cite{Courtemanche1998}, following preliminary studies that suggested it was most able to capture the biomarker values seen in the Sanchez \textit{et al.} dataset. A separate benchmarking study also suggested that this model, despite being one of the first developed for human atria, predicted APDs very well for data from both SR and cAF patients \cite{Wilhelms2013}. The CRN model uses twenty-one coupled ordinary differential equations to simulate amongst other things the activation and inactivation of nine different sarcolemmal ion channels, as well as the actions of the sarcoplasmic Ca$^{2+}$ pump, Na$^{+}$/K$^{+}$ pump and the Na$^{+}$/Ca$^{2+}$ exchanger, which all contribute to the flow of ions in or out of an atrial cell and thus to the changes in membrane potential that create the action potential. Ion channels are modelled using a Hodgkin-Huxley \cite{Hodgkin1952} type formulation, with combinations of gating variables representing the presence/absence of activators/inhibitors that determine channel availability. Ca$^{2+}$ uptake into the network sarcoplasmic reticulum, release from the junctional sarcoplasmic reticulum and the transfer (active or leak) between these two compartments are also represented. For full details of the model, including the specific forms of each of its differential equations see \cite{Courtemanche1998}.

We simulated the CRN model using MATLAB's \textit{ode15s} routine, with a maximum step size of $\Delta t = 1$ms. Biomarkers were measured from output $V(t)$ curves, after discarding any action potentials that failed to excite above $-30$mV, failed to repolarise to a value of RMP$+0.1$APA, or exhibited spontaneous depolarisations (judged as subsequent peaks after repolarisation to the aforementioned value). Following Sanchez \textit{et al.}, the model was paced until it reached steady state ($\leq$1\% change in all state variables) and then 90 more times, using a stimulus of 2ms of $-2210$ pA, approximately twice the diastolic threshold for the base model. Temperature and external ion concentrations were adjusted to match the experimental conditions ($T = 309.65$K, $[Na^+]_o = 149.42$mM, $[K^+]_o = 4.5$mM, $[Ca^{2+}]_o = 4.5$mM). No other parameters were modified from their values in the originally published version of the model.

The process of simulating the CRN model (with input parameters $\boldsymbol{\theta}$) and the subsequent calculation of biomarkers from the resulting action potential will be denoted ${\cal M}$, and the output biomarkers denoted $\mathbf{y}$, such that
\begin{equation}
\label{biomarkers}
\mathbf{y} = {\cal M}(\boldsymbol{\theta}).
\end{equation}

\subsection*{Populations of CRN Models}
Populations of models were constructed here by varying a set of inputs to the CRN model, namely the current densities of the fast Na$^{+}$ current, the five outward K$^{+}$ currents (transit outward, ultrarapid delayed rectifier, rapid and slow delayed rectifiers and the inward rectifier), the L-type inward Ca$^{2+}$ current, the Na$^{+}$/K$^{+}$ pump and Na$^{+}$/Ca$^{2+}$ exchanger, and the maximal rates of uptake and release of Ca$^{2+}$ inside the cell by the sarcoplasmic reticulum. Following the notation used in the original CRN paper, the set of inputs is here denoted $\boldsymbol{\theta} = (g_{Na}, g_{to}, g_{Kur}, g_{Kr}, g_{Ks}, g_{K1}, g_{CaL}, I_{NaK\mbox{\scriptsize(max)}}, I_{NaCa\mbox{\scriptsize(max)}}, I_{up\mbox{\scriptsize(max)}}, k_{rel})$, with $g$ denoting the maximal conductances of the different currents, $I_{\mbox{\scriptsize(max)}}$ denoting the maximal actions of pumps and exchangers and $k_{\mbox{\scriptsize rel}}$ being the conductance of the ryanodine receptors that release Ca$2^{+}$ from the sarcoplasmic reticulum. These were the same currents varied by Muszkiewicz \textit{et al.} in their construction of POMs for human atria \cite{Muszkiewicz2014}. Sanchez \textit{et al.}'s study \cite{Sanchez2014} varied only the six currents they identified as having a significant impact on biomarker values \cite{Sanchez2012}, and thus did not include $g_{Na}$, $g_{Kr}$, $g_{Ks}$, $I_{up\mbox{\scriptsize(max)}}$ or $k_{rel}$.

Following Sanchez \textit{et al.}, POMs were constructed using a search space of $\pm 100\%$ from the base parameter values for the CRN model. However, whereas their work selected trial points using the sampling method underlying Fourier amplitude sensitivity testing \cite{Marino2008} and then rejected those that did not produce APs with biomarkers falling within experimentally observed ranges, we used the method described subsequently to produce POMs that not only corresponded to the spread of the experimental data, but also reproduced its distributional features.

\subsection*{Biomarker Joint Distribution Estimation}
Distributional calibration first requires estimating the distribution represented by the data, $p(\mathbf{y})$. This was achieved here by multivariate kernel density estimation, which creates a smooth distribution by summing over a series of multivariate Gaussians centred at each of the $N$ individual datapoints,
\begin{equation}
\label{mvkde}
p(\mathbf{y}) \approx \frac{1}{\det(H)^{-1/2} N (2 \pi)^{-N_b/2}} \sum_{i=1}^{N} e^{-1/2 (\mathbf{y} - \widetilde{\mathbf{y}}_i)^T \mathbf{H}^{-1} (\mathbf{y} - \widetilde{\mathbf{y}}_i)}.  
\end{equation}
Here $N_b$ is the number of biomarkers (seven in this case), $\widetilde{\mathbf{y}_i}$ are the individual points of biomarker data and $\mathbf{H}$ is the bandwidth matrix, a parameter of the density estimator that controls the extent and direction of smoothing. When the distribution to be estimated is normal with unit variance, the optimal bandwidth can be shown \cite{Silverman1986} to be
\begin{equation}
\label{optimal_bandwidth}
h_{\mbox{\scriptsize opt}} = \left(\frac{4}{N(N_b + 2)}\right)^{\frac{2}{N_b+4}},
\end{equation}
which motivates a choice of bandwidth matrix
\begin{eqnarray}
\label{silverman_rule}
\mathbf{H}_{ij} & = & h_{\mbox{\scriptsize opt}} \, \sigma_i^2 \qquad \quad \mbox{if } i = j \\
\nonumber
 & = & 0 \qquad \qquad \quad \, \mbox{if } i \neq j 
\end{eqnarray}
Thus the extent of smoothing is weighted in each dimension in terms of the variance in that biomarker observed in the dataset, but correlations between biomarkers are ignored in the choice of $\mathbf{H}$. Note that the estimated density still attempts to account for dependencies between biomarkers. The choice to use a diagonal bandwidth matrix tends to be sufficient in practice \cite{Wand1995}.

Another alternative to multivariate kernel density estimation is to approximate $p(\mathbf{y})$ by combining the marginal distributions of each biomarker with a Gaussian copula to approximate their interdependencies. However, for the atrial datasets we use to demonstrate our calibration technique, this approach was found to be less effective and so is not discussed further here.

APD$_{20}$ readings in the SR dataset were predominantly clustered at low values, but with a considerable range. To improve the performance of the kernel density estimation (recalling the bandwidth was selected as optimal for normally distributed data), the APD$_{20}$ values were first logarithmically transformed to make their distribution more regular before use in equation (\ref{mvkde}).

\subsection*{SMC for POM Calibration to Distributions}
\label{sec:methods_dist_matching}

Constructing a population of $\boldsymbol{\theta}$ values that exhibits the estimated distribution $p(\mathbf{y})$ is not trivial, given the complex relationship between the two encoded by equation (\ref{biomarkers}). We define $g(\boldsymbol{\theta})$ to be a probability density over the space $\boldsymbol{\theta}$, and the population distribution of the model output ${\cal M}(\boldsymbol{\theta})$ when $\boldsymbol{\theta}$ is drawn according to this distribution we denote $h(\mathbf{y}|g(\boldsymbol{\theta}))$, with the vertical bar $|$ denoting conditioning. That is, $h(\mathbf{y}|g(\boldsymbol{\theta}))$ is the density of ${\cal M}(\boldsymbol{\theta})$ when $\boldsymbol{\theta}\sim g(\boldsymbol{\theta})$:
\begin{equation}
\nonumber
h(\mathbf{y}|g(\boldsymbol{\theta})) = \lim_{\Delta_{\mathbf{y}} \rightarrow \mathbf{0}} \frac{1}{\prod_{k=1}^{N_b}\Delta_{b_k}} \int_{{\cal M}(\boldsymbol{\theta}) \in (\mathbf{y}, \mathbf{y} + \Delta_{\mathbf{y}})} g(\boldsymbol{\theta}) d\boldsymbol{\theta},
\end{equation}
where $\Delta_{\mathbf{y}} = (\Delta_{b_1},\ldots,\Delta_{b_{N_b}})$. Our problem is thus recast as finding the distribution $g(\boldsymbol{\theta})$ that produces a $h(\mathbf{y}|g(\boldsymbol{\theta}))$ that is as close as possible to $p(\mathbf{y})$. Models sampled according to this optimal $g(\boldsymbol{\theta})$ will then exhibit outputs that reproduce the estimated distribution of outputs in the data. 

If the ``closeness'' of $h(\mathbf{y}|g(\boldsymbol{\theta}))$ and $p(\mathbf{y})$ is measured in terms of the Kullback-Leibler divergence between the two distributions, the problem is an optimisation problem
\begin{equation}
\label{kld}
g^*(\boldsymbol{\theta}) = \arg \min_{g(\boldsymbol{\theta})} \int_{\mathbf{y}} \ln \left( \frac{h(\mathbf{y}|g(\boldsymbol{\theta}))}{p(\mathbf{y})}  \right) h(\mathbf{y}|g(\boldsymbol{\theta})) d\mathbf{y}.
\end{equation}
The standard method for solving this type of problem is to follow a variational Bayes (VB) type of approach.  VB is commonly used to produce parameteric approximations of posterior distributions in Bayesian statistics \cite[chap.\ 11]{Bishop2006}.  Applying VB to our problem would involve specifying some parametric distribution for $g(\boldsymbol{\theta}) \equiv g(\boldsymbol{\theta}|\boldsymbol{\phi})$, where $\boldsymbol{\phi}$ are the parameters of the distribution.  For example, if a multivariate normal distribution was adopted for $g$, $\boldsymbol{\phi}$ would consist of a mean vector and covariance matrix. Using VB, the problem (\ref{kld}) then reduces to finding the optimal parameters $\boldsymbol{\phi}^* = \arg \min_{\boldsymbol{\phi}}f(\boldsymbol{\phi})$, with the integral in equation (\ref{kld}) approximated by Monte Carlo integration taking $K$ independent draws from $g(\boldsymbol{\theta}|\boldsymbol{\phi})$,
\begin{equation}
\nonumber
\int_{\mathbf{y}} \ln \left( \frac{h(\mathbf{y}|g(\boldsymbol{\theta}|\boldsymbol{\phi}))}{p(\mathbf{y})}  \right) h(\mathbf{y}|g(\boldsymbol{\theta}|\boldsymbol{\phi})) d\mathbf{y} \approx \frac{1}{K}\sum_{k=1}^{K} \ln \left( \frac{\hat{h}(\mathbf{y}_k|g(\boldsymbol{\theta}|\boldsymbol{\phi}))}{p(\mathbf{y}_k)}  \right) = f(\boldsymbol{\phi}),
\end{equation} 
where $\mathbf{y}_k = {\cal M}(\boldsymbol{\theta}_k)$ and $\boldsymbol{\theta}_k \sim g(\boldsymbol{\theta}|\boldsymbol{\phi})$ for $k=1,\ldots,K$.  For a particular $\boldsymbol{\phi}$, an estimate of $h(\mathbf{y}|g(\boldsymbol{\theta}|\boldsymbol{\phi}))$, which we denote as $\hat{h}(\mathbf{y}_k|g(\boldsymbol{\theta}|\boldsymbol{\phi}))$, could be obtained using a kernel density estimate as in equation (\ref{mvkde}) based on the set of simulated biomarker values $\{\mathbf{y}_k\}_{k=1}^K$. There are several reasons why we did not adopt this approach: firstly it requires us to specify a parametric form for $g(\boldsymbol{\theta})$, secondly evaluating $f(\boldsymbol{\phi})$ is very expensive as it involves solving the model $K$ times and thirdly $\boldsymbol{\phi}$ will be high-dimensional, leading to a difficult optimisation problem. 

Instead, we used an approach that is more pragmatic and effective in this application.  First, we determined a collection of $\boldsymbol{\theta}$ values (or models) that produced biomarker values $\mathbf{y}$ that have relatively high density with respect to the data density $p(\mathbf{y})$. Then, we removed models from this collection in an iterative fashion so that the distribution of corresponding biomarker values that remained was even closer to $p(\mathbf{y})$ (see Section ``Further POMs Refinement'').

For the first step we used sequential Monte Carlo (SMC, \cite{DelMoral2006}), following the use of the technique to construct POMs calibrated to ranges in data \cite{Drovandi2016}. This technique begins with a set of $N$ particles and traverses them through a sequence of probability distributions by iteratively applying importance sampling, resampling and move steps. We achieve this behaviour by sampling from the sequence of distributions $h(\boldsymbol{\theta}) \propto p(\mathcal{M}(\boldsymbol{\theta}))^{\gamma}$ , with $\gamma \in [0,1]$. It can immediately be seen that $\gamma = 0$ corresponds to the uniform distribution, which is very easy to sample if we specify some lower and upper limits for each component of $\boldsymbol{\theta}$, and that $\gamma = 1$ corresponds to a distribution proportional to $p(\mathcal{M}(\boldsymbol{\theta}))$, which is potentially difficult to sample. Successively incrementing $\gamma$ after each resample and move step until $\gamma$ reaches one allows for the complexity of the sampling problem to be introduced gradually. An important aspect of SMC is that it does not require the distributions in the sequence to be properly normalised. The full SMC algorithm is laid out in the supplementary material.

The algorithm requires the use of traditional Markov chain Monte Carlo (MCMC, \cite{Brooks2011}) steps in order to find unique locations for particles after each resampling step, and the particles must still represent samples from the current target distribution. This is achieved using the Metropolis-Hastings algorithm, which, in our set-up, accepted or rejected any proposed particle moves according to
\begin{equation}
\label{metropolis_hastings_model}
\mbox{Pr}(\mbox{accept}) = \min \left( 1, \frac{[p({\cal M}(\boldsymbol{\theta}_{\mbox{\scriptsize new}}))]^{\gamma}{\cal J}(\boldsymbol{\theta}_{\mbox{\scriptsize old}}|\boldsymbol{\theta}_{\mbox{\scriptsize new}})}{[p({\cal M}(\boldsymbol{\theta}_{\mbox{\scriptsize old}}))]^{\gamma}{\cal J}(\boldsymbol{\theta}_{\mbox{\scriptsize new}}|\boldsymbol{\theta}_{\mbox{\scriptsize old}})} \right).
\end{equation}
Here ${\cal J}$ is the jumping distribution that generates proposed moves of particles, and in our case does not depend on the previous particle location (that is, ${\cal J}(\boldsymbol{\theta}_{\mbox{\scriptsize new}}|\boldsymbol{\theta}_{\mbox{\scriptsize old}}) = {\cal J}(\boldsymbol{\theta}_{\mbox{\scriptsize new}})$). The jumping distribution used is a Gaussian mixture model built for a regularised version of the current of locations of all particles after a resampling step (see SMC Algorithm in the supplementary material). The number of MCMC steps performed after each resampling step is chosen adaptively \cite{Drovandi2011}.

The output of the SMC algorithm is a collection of samples $\{\boldsymbol{\theta}_i\}_{i=1}^N$ from a distribution proportional to $p(\mathcal{M}(\boldsymbol{\theta}))$.  We note that the corresponding collection of output values $\{\mathbf{y}_i\}_{i=1}^N$, where $\mathbf{y}_i = \mathcal{M}(\boldsymbol{\theta}_i)$, is not a sample from the distribution of interest, $p(\mathbf{y})$.  The reason for this is that we have not accounted for the nonlinear transformation, $\mathbf{y} = \mathcal{M}(\boldsymbol{\theta})$ to correctly convert the target distribution over the output space to the corresponding target distribution over the parameter space. However, given the fact that the transformation $\mathbf{y} = \mathcal{M}(\boldsymbol{\theta})$ is not analytic and not a one to one function, we suggest that it is not tractable to properly account for it and indeed it may not even be possible to find a distribution over $\boldsymbol{\theta}$ that leads to a distribution of biomarker values consistent with $p(\mathbf{y})$.

Nonetheless, we found that this approach led to a collection of parameter values that generated biomarker values with relatively high density under $p(\mathbf{y})$.  We used this collection as the starting point for our subsequent refinement process.

As an alternative to SMC, an MCMC approach could also be used directly to produce samples from a distribution proportional to $p(\mathcal{M}(\boldsymbol{\theta}))$, using the same density (\ref{mvkde}) and acceptance algorithm (\ref{metropolis_hastings_model}) with $\gamma = 1$. Indeed we used a modern state-of-the-art MCMC sampler, DiffeRential Evolutionary Adaptive Metropolis (DREAM) \cite{Vrugt2016} to verify our SMC algorithm and found that the two produced comparable results in terms of the distribution of the values of $\boldsymbol{\theta}$ produced. The primary benefit of the SMC algorithm, when it came to the construction of our initial POM, was that the output of the algorithm is a set of unique samples from the distribution almost the same size as the number of particles, $N$, which is specified by the user. In contrast, MCMC approaches must be run an indefinite amount of time until the chain has been judged to have converged, producing a long chain of samples of initially unknown length. Moreover, these chains contain many repeated samples, and filtering out these repeats will also destroy the desired distribution.

\subsection*{Further POMs Refinement}
\label{sec:methods_refinement}
The approximate nature of the SMC calibration process encouraged further refinement of the constructed POMs in order to fully capture the statistical distributions seen in the data. This was achieved by selecting a subset of the population such that the new smaller set of models better exhibited the biomarker distributions seen in the data. In order to do this, first a quantitative measure of how well a POM captured the distributions observed in the data was constructed. We used the Jensen-Shannon distance ($JSD$), a symmetric and finite version of the Kullback-Leibler divergence that remains a measure of the `distance' between two probability distributions. Labelling the two distributions $p(\mathbf{y})$ and $q(\mathbf{y})$, the Jensen-Shannon distance is given by
\begin{equation}
\label{JSD}
JSD = \Biggl[ \frac{1}{2} \int_{\mathbf{y}} p(\mathbf{y}) \ln \left( \frac{p(\mathbf{y})}{\frac{1}{2} p(\mathbf{y}) + \frac{1}{2} q(\mathbf{y})} \right) \, d\mathbf{y} + \frac{1}{2} \int_{\mathbf{y}} q(\mathbf{y}) \ln \left( \frac{\mathbf{y})}{\frac{1}{2} p(\mathbf{y}) + \frac{1}{2} q(\mathbf{y})} \right) \, d\mathbf{y} \Biggr]^{1/2},
\end{equation}
with the square root used to make the divergence measure a metric \cite{Endres2003}.

When the data is high-dimensional (say $N \geq 5$), the `full' JSD between the multivariate joint distributions of the observations in the dataset and those generated by a given POM is difficult to calculate accurately. Therefore, we instead used the distances between the marginal distributions for each of the biomarkers, $JSD_i$, along with the distances between the bivariate distributions between all possible pairs of observation variables, $JSD_{ij}$ to create a matrix, the norm of which serves as an approximate measure of fit, namely
\begin{align}
\label{rho}
\rho &= \left|\left| \mathbf{P} \right|\right|_2, \\
\nonumber
\mathbf{P}_{ij} &= \left\{ \begin{array}{lr} \! \! \! JSD_{i} & \quad i = j \\
\! \! \! JSD_{ij} & \quad i \neq j \end{array}\right. \qquad \qquad (i, j = 1,...,N_B).
\end{align}
The measure $\rho$ takes into account how well the individual distributions of each observed variable are represented by a POM, along with some measure of how well it captures the dependency between these variables. We note that this is not necessarily the best measure of fit, but uses more easily calculated divergences to produce a single value, allowing the use of the technique described below. JSD values used in the calculation of $\rho$, as integrals, were approximated using standard Riemann integration.

Our refinement process seeks to minimise $\rho$, using a supplied POM. In this work we use POMs selected using SMC to improve fit with the data, though we note that POMs constructed using typical Monte Carlo sampling techniques (such as LHS and calibrated to the ranges of the data) could also be used as starting points for our refinement procedure. Minimisation is achieved by trialling removal of individual models from the population (or re-introduction of removed models) and then accepting or rejecting them according to the Metropolis probability,
\begin{equation}
\label{metropolis_annealing}
\mbox{Pr}(\mbox{accept}) = e^{-\Delta\rho/T}.
\end{equation}
Here $\Delta \rho$ is the change in the overall divergence measure (\ref{rho}) associated with the trialled removal/re-introduction and $T$ is a parameter of the process that controls the likelihood of accepting unfavourable trial updates. This approach is very similar to the approach of simulated annealing \cite{Kirkpatrick1983}, although we use a fixed value of $T = 0.1$ instead of gradually decreasing it. Every 1000 trial steps, we judge if the choice of subpopulation is wandering too far away from the optimum by checking if $\rho$ is more than 1\% larger than the current best $\rho$ value found, and if so, restart the process back to the configuration corresponding to the best $\rho$ value.

The only additional condition we use is that the size of the subpopulatoin of models cannot fall below the number of datapoints, ensuring that the resulting population will not become small enough to lose meaning. If a larger population of models is desired, $\rho$ (representing the `energy' of the system that is minimised over the course of the annealing process) can be replaced by a new expression in equation (\ref{metropolis_annealing}) that penalises both higher values of $\rho$ and small numbers of models in the population.

We saw (see Results) that the data for one biomarker, namely the maximum upstroke velocity, took values not predicted by the CRN model in the search space, and strong correlations exhibited by the model were not seen in the data. This made it appealing to de-emphasise the contributions of this biomarker to the POM refinement process. This was achieved by creating a second divergence measure, $\hat{\rho}$, that is the 2-norm of a modified version of the performance matrix $\mathbf{P}$ with the row and column corresponding to the maximum upstroke velocity overwritten with zeroes, except for the diagonal element. Minimising $\hat{\rho}$ instead allowed the distributions of the other biomarkers to be better fit by the refinement process, at the cost of producing POMs that did not strongly reflect the distribution of maximum upstroke velocities in the data. Given that we expected the maximum upstroke velocity to be the biomarker most subject to experimental measurement error in the data, we consider this a reasonable decision.

\section*{Acknowledgements and Funding}

The authors would like to thank Dr. Xin Zhou and Anna Muszkiewicz of the University of Oxford, and Dr. Carlos Sanchez of the University of Zaragoza for ongoing discussions that contributed to this work, and Dr. Carlos Sanchez for providing the atrial biomarker data (available in online supplement for Sanchez \textit{et al.} 2014). We would also like to thank Giuseppe Di Martino for performing initial studies. Facilities used to simulate the computational model used in this work were provided by the HPC and Research Support Group, Queensland University of Technology, Brisbane, Australia. All authors contributed to the drafting and revision of the manuscript. BL, CD, PB, BR and KB contributed to the design of experiments. BL, CD and NC performed experiments. BL, CD, PB, BR and KB contributed to the analysis of results. No authors have competing interests. BR, PB and KB are funded by the Australian Research Council under grant number CE-140100049. CD is funded by the Australian Research Council's Discovery Early Career Researcher Award scheme under grant number DE-160100741. NC is supported by the Basque government's BERC 2014-2017 program, the Spanish Ministry of Economy and Competitiveness through BCAM Severo Ochoa excellence accreditation SEV-2013-0323 and through project MTM2015-69992-R "BELEMET". BR is funded by a Wellcome Trust Senior Research Fellowship in Basic Biomedical Science (100246/Z/ 12/Z), the British Heart Foundation Centre of Research Excellence in Oxford (RE/13/1/30181), an NC3R Infrastructure for Impact award (NC/P001076/1), an EPSRC Impact Acceleration Award (EP/K503769/1), the ComBioMed project funded by the European Union’s Horizon 2020 research and innovation programme (grant agreement \#675451)."

\bibliographystyle{ieeetr}
\bibliography{matching_dists}

\clearpage

\section*{Supplementary Materials}
\beginsupplement

\begin{table}[ht]
\begin{center}
\begin{tabular}{lllll}
\hline
\bf{Biomarker}  & \bf{Range} & \bf{Mean} & \bf{Std. Dev.} & \bf{JSD} \\ \hline
APD$_{90}$ (ms) & 148$ - $351 (141$ - $349) & 216 (216) & 33 (35) & 0.032 \\
APD$_{50}$ (ms) & 33$ - $168 (36$ - $182) & 101 (102) & 28 (28) & 0.040 \\
APD$_{20}$ (ms) & 2$ - $114 (4$ - $82) & 29 (30) & 20 (18) &  0.075 \\
APA (mV) & 81$ - $119 (74$ - $133) & 102 (102) & 7 (8) &  0.075 \\
RMP (mV) & -87$ - $-68 (-90$ - $-66) & -77 (-77) & 3 (4) & 0.073  \\
V$_{20}$ (mV) & -29$ - $21 (-33$ - $21) & -4 (-4) & 10 (11) & 0.049 \\
$\de{V}{t}_{\mbox{\scriptsize max}}$ (V/s) & 101$ - $301 (40 $ - $414) & 189 (232) & 34 (70) & 0.361  \\ \hline
\end{tabular}
\end{center}
\caption{{\bf{Summary statistics for the cAF dataset are well recovered by the calibrated POM.}} Summary statistics for the POM calibrated to the distributions in biomarkers exhibited by atrial cells from hearts exhibiting cAF, as compared to the summary statistics for the experimental data itself (given in parentheses). Deviation in the marginal distributions of each biomarker are specified in terms of the Jensen-Shannon distance (JSD), calculated using equation (\ref{JSD}). As with the case of the SR data, the statistical distribution is well captured apart from  the maximum upstroke velocity. Values shown are for the POM obtained by minimising $\hat{\rho}$.}
\label{tab:CA_summaries}
\end{table}

\clearpage

\begin{algorithm}
\caption{SMC algorithm for construction of a POM fitted to an underlying distribution of biomarkers, $p(\mathbf{y})$.}
\begin{algorithmic}[0]
\State
\LineComment{Initialise particles}
\State Set $i = 0$
\While{$i \leq N_{\mbox{\scriptsize parts}}$}
\State Select a random $\boldsymbol{\theta}$ from the search space and calculate $\mathbf{y} = {\cal M}(\boldsymbol{\theta})$
\If{action potential not rejected (see Materials and Methods)}
\State Set $i = i+1$
\State Store particle location in parameter space, $\boldsymbol{\theta}_i$ and biomarkers, $\mathbf{y}_i$
\State Store particle likelihood, ${\cal L}_i = p(\mathbf{y_i})$
\EndIf
\EndWhile
\State
\LineComment{Gradually increment $\gamma$ until the true distribution is sampled}
\State Set $\gamma = 0$
\While{$\gamma < 1$}
\State
\LineComment{Check if current particles sufficiently reproduce the desired distribution}
\If{ESS($\gamma, 1) \geq \:\! N_{\mbox{\scriptsize parts}}/2$}
\State Set $\gamma = 1$
\Else
\State Find $\gamma^{\prime}$ such that ESS($\gamma, \gamma^{\prime}$) $= \:\! N_{\mbox{\scriptsize parts}}/2$
\EndIf
\State
\LineComment{Resample particles according to the new distribution}
\State Calculate normalised weights for particles, $w_i = {\cal L}_i\:\!^{(\gamma^{\prime} - \gamma)} / \sum_{j=1}^{N_{\mbox{\scriptsize parts}}}{\cal L}_j\:\!^{(\gamma^{\prime} - \gamma)}$
\State Resample particle locations $\boldsymbol{\theta} \sim \mbox{Multinomial}(w)$
\State Update $\gamma \rightarrow \gamma^{\prime}$
\State
\LineComment{Attempt to remove particle duplications via MCMC move steps}
\State Construct the jumping distribution, ${\cal J}(\boldsymbol{\theta}) =$ \Call{BuildJumpDist}{$\boldsymbol{\theta}$}
\State Update particle locations, $[\boldsymbol{\theta},\mathbf{y},\mbox{acc}]$ = \Call{MCMCMove}{$\boldsymbol{\theta},\mathbf{y}$}
\State Determine optimal number of MCMC iterations, $R = \mathrm{ceil}\left(\frac{\ln 0.05}{\ln(1 -\mathrm{acc})}\right)$
\For{$i = 1$ to min($R-1,29$)}
\State $[\boldsymbol{\theta}, \mathbf{y},\sim] = $ \Call{MCMCMove}{$\boldsymbol{\theta},\mathbf{y},{\cal J}(\boldsymbol{\theta})$}
\EndFor
\State
\EndWhile
\end{algorithmic}
\label{alg:smc}
\end{algorithm}

\begin{algorithm}
\caption{Ancillary functions used by the SMC algorithm}
\begin{algorithmic}[0]
\State
\Function{ESS}{$\gamma,\gamma^{\prime}$}
\State Calculate particle weights, $w_i = {\cal L}_i\:\!^{(\gamma^{\prime} - \gamma)} / \sum_{j=1}^{N_{\mbox{\scriptsize parts}}}{\cal L}\:\!^{(\gamma^{\prime} - \gamma)}$
\State Return estimated sample size, ESS $= 1 / \sum_{j=1}^{N_{\mbox{\scriptsize parts}}}w_j\:\!^2$
\EndFunction
\State
\Function{${\cal J}(\boldsymbol{\theta}) = $ BuildJumpDist}{$\boldsymbol{\theta}$}
\LineComment{Regularise the marginal distributions of $\boldsymbol{\theta}$}
\State Scale particle locations to $[0,1]$, $\boldsymbol{\phi}_i = \frac{\boldsymbol{\theta}_i - \boldsymbol{\theta}_{\mbox{\scriptsize min}}}{\boldsymbol{\theta}_{\mbox{\scriptsize max}} - \boldsymbol{\theta}_{\mbox{\scriptsize min}}}$
\State Fit a beta distribution to the values of $\boldsymbol{\phi}$.
\State Use this to find an optimal mixture of two beta distributions, $f(\boldsymbol{\phi})$
\State Use the cdf of the beta mixture, $\mathbf{u}_i = F(\boldsymbol{\phi}_i)$ to obtain approximately uniformly distributed particles
\State Transform these into normally distributed particles, $\mathbf{z} = \mbox{norminv}(\mathbf{u})$
\LineComment{Jumping dist. is Gaussian mixture model on regularised distributions}
\State Fit a mixture of three Gaussians to particle $\mathbf{z}$'s using MATLAB's \textit{fitgmdist}
\State Store the Gaussian mixture model, ${\cal J}(\mathbf{z})$
\State Calculate and store ${\cal J}(\mathbf{z})$ for all particles
\EndFunction
\State
\Function{[$\boldsymbol{\theta}, \mathbf{y}] = $MCMCMove}{$\boldsymbol{\theta},\mathbf{y},{\cal J}(\boldsymbol{\theta})$}
\For{$i = 1$ to $N_{\mbox{\scriptsize parts}}$}
\State Propose $\mathbf{z}_i^{\prime} \sim {\cal J}(\mathbf{z})$
\State Transform $\mathbf{z}_i^{\prime}$ back to $\boldsymbol{\theta}_i^{\prime}$
\State Evaluate the model, $\mathbf{y}_i^{\prime} = {\cal M}(\boldsymbol{\theta}_i^{\prime})$
\LineComment{Accept or reject according to Metropolis-Hastings algorithm}
\State Generate a uniform random number $r \sim [0,1]$
\If{$r < \min \left( 1, \frac{[p(\mathbf{y}_i^{\prime})]^{\gamma}{\cal J}(\boldsymbol{\theta}_i)}{[p(\mathbf{y}_i)]^{\gamma}{\cal J}(\boldsymbol{\theta}_i^{\prime})} \right)$}
\State Update $\boldsymbol{\theta}_i \rightarrow \boldsymbol{\theta}_i^{\prime}$, $\mathbf{y}_i \rightarrow \mathbf{y}_i^{\prime}$
\EndIf
\EndFor
\EndFunction
\end{algorithmic}
\label{alg:ancillary}
\end{algorithm}

\end{document}